\documentstyle[12pt]{article}

\input epsf

\def\setb@se#1{\baselineskip=#1 \normalbaselineskip=#1}
\setb@se{14pt}
\newcommand{\be}{\begin{equation}}
\newcommand{\ee}{\end{equation}}
\newcommand{\g}{{\bf g}}
\newcommand{\x}{{\bf x}}
\newcommand{\M}{{\bf M}}
\newcommand{\D}{{\bf D}}
\newcommand{\A}{{\bf a}}

\newcommand{\mm}{\mu_{0}}

\newcommand{\La}{{\bf L}}
\newcommand{\Ta}{{\bf T}}
\newcommand{\Sa}{{\bf S}}
\newcommand{\Ja}{{\bf J}}
\newcommand{\Ka}{{\bf K}}
\newcommand{\newp}{  }
\renewcommand{\theequation}{\arabic{section}.\arabic{equation}}
\begin{document}

\begin{titlepage}
\begin{flushright}
ZU-TH-11/96

hep-th/9604054
\end{flushright}

\vspace{20 mm}
\begin{center}
{\Huge \bf Computation of the winding number
diffusion rate due to the  cosmological sphaleron\\
}

\vspace{10 mm}
\end{center}
\begin{center}
{\bf  Mikhail S. Volkov}

\vskip2mm
{\em Institut f\"ur Theoretische Physik der
Universit\"at Z\"urich--Irchel, \\
Winterthurerstrasse 190, CH--8057 Z\"urich,
Switzerland,\\
e--mail: volkov@physik.unizh.ch}
\end{center}

\vspace{20mm}\begin{center}{\bf Abstract}\end{center}
\vspace{2mm}

\noindent
A detailed quantitative analysis of the
transition process mediated by a sphaleron type  
non-Abelian gauge field configuration in a static Einstein universe
is carried out. By examining spectra of the fluctuation operators
and applying the zeta function regularization scheme,
a closed analytical expression for
the transition rate at the one-loop level is derived. 
This is a unique example of an exact solution for a
sphaleron model in $3+1$ spacetime dimensions.

\vspace{5 cm}

\noindent

\end{titlepage}

\section{Introduction}

The discovery of vacuum periodicity \cite{rebbi}, \cite{thooft} and
sphalerons \cite{Manton} in the  standard
model has revealed a considerable fermion number 
violation in the theory at high temperatures.
The fermion number changes in the processes of barrier transitions
between the distinct topological sectors of the theory,
where the barrier height is determined by the sphaleron energy. 
Such processes become
unsuppressed at temperatures of the order
of the sphaleron mass.
Since the different topological sectors 
are in thermal equilibrium at high
temperatures, the sphaleron transitions lead to the dissipation
of the baryon asymmetry produced at earlier times,
that is, during the electroweak phase transition or at GUT energies
\cite{kuzmin}.
The estimates for the transition rate \cite{arnold}-\cite{diak}
show that the sphaleron mechanism can be efficient enough
to reduce the asymmetry by several orders of magnitude.

The sphaleron solution in the standard model is known only
numerically,
which causes  considerable computational difficulties
\cite{carson}-\cite{diak}. 
Other field-theoretical sphaleron models
have therefore become important \cite{Forgacs},
\cite{bochkarev}, \cite{mottola}.
Some of these models are exactly solvable
in the sense that the corresponding sphaleron transition rate can be
evaluated analytically at the one-loop level
\cite{bochkarev}, \cite{mottola}, which provides
a closer insight into the physics. Unfortunately,
all of these
models exist only in low spacetime dimensions.
There is also  an alternative example of four dimensional
sphalerons which appear in the theory of a self-gravitating
non-Abelian gauge field.
Gravity violates the scale invariance and plays therefore a role
similar to that
of a Higgs field. This results in the existence of classical \cite{BK}
sphaleron-like solutions in the theory
\cite{10}, \cite{12}, which are also known only numerically.

The purpose of the present paper is to investigate a
sphaleron model that is exactly solvable and yet exists in
$3+1$ spacetime dimensions, which distinguishes it from the 
other known models. Similarly to the example mentioned above, 
this model deals with
a non-Abelian gauge field interacting with gravity,
but now gravity is regarded as a fixed external field.
Specifically, we shall consider the theory of
a pure $SU(2)$ Yang-Mills field in a static Einstein universe.
This theory admits an analytically known sphaleron solution
which we shall call cosmological sphaleron.  
The analysis of the corresponding transition problem 
is the subject of this paper.

In fact, this type of sphaleron was found long ago.
The solution itself has been discussed in various contexts,   
together with the other related solutions of the
(Einstein)-Yang-Mills
field equations \cite{Hosotani}-\cite{cosmEYM}. 
The sphaleron nature of the solution was also
recognized and discussed in connection with the finite volume
QCD \cite{baal} and  in a cosmological context as well
\cite{cosmsphal}, \cite{ding}.
However, an analysis of the corresponding sphaleron transition
problem has been lacking so far. 

The cosmological sphaleron is
distinguished by the important property that the sphaleron
configuration consists of the gauge field alone. This
implies that the sphaleron is not plagued with the various
symmetry breaking phase transitions, and this also ensures that the
dynamics of the sphaleron mediated processes is
conformally invariant --
up to the anomalous scale dependence of the gauge coupling
constant. Since the sphaleron has very high 
symmetry, there is an opportunity 
to analyze the sphaleron transition
problem analytically at the one-loop level, 
which is of great methodological interest. 
(We shall use the names ``cosmological'' and 
``universe'', although the model can be applied not only
in the cosmological context.)

The problem which will be investigated below
can be formulated as follows.
Consider a gauge field in a static Einstein universe
at finite temperature.
Specifically, consider the thermal ensemble over one
of the topological vacua of the field.
Find the rate of the decay of this thermal state due to the
diffusion into the neighboring topological sector. In order
to obtain the answer, we use the Langer-Affleck formula
and take only the bosonic degrees of freedom into account
(the fermion contribution can be considered in a similar way).
We do not assume the high temperature limit and take the sum over
all Matsubara modes. We use the zeta function
regularization scheme, which allows us to entirely carry out the
analysis.
Our principal results are given by Eqs.(\ref{5:20a})-(\ref{5:24}) and
Eq.(\ref{5:30}) and presented in Figs.1-3.
The rest of the paper is organized as follows. The basic
properties of the model, such as topological vacua and
the sphaleron solution are discussed in Sec.II.
A brief derivation, based on
the path integral methods, of the decay rate 
of an unstable phase at finite temperature is given in Sec.III. 
The path integration procedure is outlined in Sec.IV.
The spectra of the fluctuation operators are analyzed in Sec.V.
The determinants of these operators are calculated
in Sec.VI which also includes the
consideration of the high temperature limit. Sec.VII. contains some
concluding remarks, and the derivation of numerous
formulae used in the zeta function approach is given
in the Appendix.

Throughout the paper the units $\hbar=c=\kappa_{B}=1$ are used.
The symbol $g$ stands for the gauge coupling constant, whereas
the spacetime metric is denoted by \g.

\newp
\section{The sphaleron on $S^{3}$}
\setcounter{equation}{0}

Consider the static Einstein universe $(M,\g)$,
where $M=R^{1}\times S^{3}$, and the metric is 
\be
ds^{2}=\A^{2}(-d\eta^{2}+d\Omega^{2}_{3}).              \label{1}
\ee
Here $\A$ is a constant scale factor,
and the line element on $S^{3}$ is parameterized by
\be
d\Omega^{2}_{3}=d\xi^{2}+sin^{2}\xi
(d\vartheta^{2}+sin^{2}\vartheta d\varphi^{2}),          \label{2}
\ee
where $\xi\in [0,\pi]$, and $\vartheta$, $\varphi$ are the
usual spherical coordinates on $S^{2}$. 

The model under consideration is defined by the action
\be                                                     \label{2:a}
S[A]=-\frac{1}{2g^{2}}tr\int_{M} F_{\mu\nu}F^{\mu\nu}
\ \sqrt{-\g}d^{4}x,                                           
\ee
where $A=A_{\mu}dx^{\mu}=
T_{p}A^{p}_{\mu}dx^{\mu}$ is the gauge field,  and
$F_{\mu\nu}=\partial_{\mu}A_{\nu}-\partial_{\nu}A_{\mu}-
i[A_{\mu},A_{\nu}]$ is the field tensor.  
The hermitian group generators are
$T_{p}=\tau^{p}/2$  with $\tau^{a}$ being the Pauli matrices.
The  metric $\g$ in Eq.(\ref{2:a}) is given by Eq.(\ref{1}),
such that the action defines the theory of a pure non-Abelian $SU(2)$
gauge field in the static Einstein universe. (Throughout the paper 
we neglect the back reaction of the gauge field on the spacetime
geometry, which can be justified if $\A\gg l_{pl}/g$,
where $l_{pl}$ is Planck's length.)
The classical equations of motion following from the action are
\be                                                                            
               \label{2:b}
\nabla_{\mu}F^{\mu\nu}-i[A_{\mu},F^{\mu\nu}]=0,
\ee
where $\nabla_{\mu}$ is the covariant derivative with respect to the
spacetime metric.
 
First, we need to describe the topological vacua of the
gauge field in this case. We do so by
introducing a smooth, time-independent
function on the manifold, $U(\x)\in SU(2)$, where
$\x=x^{m}$, $m=1,2,3$,
thus defining the mapping
\be
U(\x):\ \ \ S^{3}\rightarrow SU(2).                 \label{3}
\ee
Any such mapping can be
characterized by an integer winding number
\be
{\bf k}[U]=\frac{1}{24\pi^{2}}\ tr\int_{S^{3}}
UdU^{-1}\wedge UdU^{-1}\wedge UdU^{-1},                \label{4}
\ee
such that the set of all $U$'s falls into
a countable sequence of  disjoint homotopy classes.
The representative of the $k$-th class, $U^{(k)}$,
${\bf k}[U^{(k)}]=k$, can be chosen as
\be                                                 \label{5}
U^{(k)}(\x)=U^{(k)}(\xi,\vartheta,\varphi)=
\exp\left\{- i k\xi\ n^{a}\tau^{a}\right\},                     
\ee
where $n^{a}=(\sin\vartheta\cos\varphi, \sin\vartheta\sin\varphi,
\cos\vartheta)$. 
The functions $U$ generate static gauge transformations, 
\be
A\rightarrow UAU^{-1}+iUdU^{-1}.                       \label{5:a}
\ee
The elements of the zeroth homotopy class, $U^{(0)}$, 
give rise to 
small gauge transformations which can be continuously
deformed to identity; other functions, $U^{(k)}$, $k\neq 0$,
generate  large transformations. A vacuum of the gauge 
field is a pure gauge, $A_{vac}=iUdU^{-1}$. Since all $U$'s
split into homotopy classes, all pure gauges
decompose into disjoint sets called topological vacua. 
The $k$-th topological vacuum in the temporal gauge is
\be
A^{(k)}(\x)=iU^{(k)}dU^{(k)-1}.                          \label{6}
\ee
By construction, the Chern-Simons number of this field 
configuration coincides with the winding number $k$.

Distinct topological vacua can not be joined by a continuous
interpolating sequence of pure gauge configurations, $iUdU^{-1}$,
since this would require a change in the winding number
${\bf k}[U]$. However, one can join them by a family of
non-vacuum fields.
That is how one can see
that the model admits a sphaleron solution.
Consider the two neighboring vacua given by Eqs.(\ref{5}),
(\ref{6}): $A^{(0)}=0$ and $A^{(1)}$.
They can by joined by the following path in the configuration space:
\be
A[h]=i\frac{1+h}{2}U^{(1)}dU^{(1)-1},               \label{7}
\ee
where the parameter $h\in [-1,1]$. (Applying a large gauge
transformation one can reduce such a path to a non contractible
loop.)  The energy-momentum tensor for this field is
\be                                                                            
               \label{10:1}
T^{\mu}_{\nu}(A[h])=
\frac{(h^{2}-1)^{2}}{2g^{2}\A^{4}}{\rm diag}(3,-1,-1,-1),  
\ee
such that the energy is given by
\be
E[h]=\int T^{0}_{0}\sqrt{^{3}\g}d^{3}x=
\frac{3\pi^{2}}{g^{2}\A}(h^{2}-1)^{2}.            \label{8}
\ee
This function
has the typical barrier shape: it vanishes at the vacuum
values of $h$, $h=\pm 1$, and reaches its maximum in between,
at $h=0$. The top of the barrier relates to the
field configuration
\be
A^{(sp)}\equiv A[h=0]=
\frac{i}{2}\ U^{(1)}dU^{(1)-1},                \label{8:1}
\ee
with the energy $E_{max}=3\pi^{2}/g^{2}\A$. 

Similarly, one can
define $E_{max}$ for any other interpolating path,
and then minimize the result over all paths.
If a non zero minimum exists then it relates to
an unstable classical solution called sphaleron.
By construction, 
the sphaleron energy defines the minimal height of the
potential barrier \cite{Manton}, \cite{Forgacs}.

To carry out such a program would, however,  be too difficult
a task which has never been done in reality. Instead, we simply
check that the field (\ref{8:1}) solves the classical equations of motion.
To see this, it is illuminating to allow the parameter $h$ in Eq.(\ref{7})
to depend on time, $h\rightarrow h(\eta)$. 
Then the Yang-Mills equations (\ref{2:b}) for the field (\ref{7})
reduce to one non-trivial equation,
which admits a first integral:
\be                                                      \label{9}
\frac{d^{2}h}{d\eta^{2}}+2h(h^{2}-1)=0,\ \ \ \ \Rightarrow \ \ \ \ \
\left(\frac{dh}{d\eta}\right)^{2}+(h^{2}-1)^{2}=\varepsilon,  
\ee
with $\varepsilon$ being an integration constant. Effectively,
these equations describe a particle moving in the one-dimensional
double-well potential. When $\varepsilon=1$, one finds
the static solution $h(\eta)=0$, which describes an unstable
equilibrium of the particle on the top of the barrier. 
This shows that  field configuration (\ref{8:1})
indeed solves the equations of motion and
relates to a saddle point of the energy functional, such that 
it can be naturally called sphaleron. Later we shall
see that this solution has only one unstable mode.
In addition,
for $\varepsilon=1$, Eqs.(\ref{9}) admit the solution
for the particle rolling down the barrier,
$h(\eta)=\sqrt{2}/\cosh \sqrt{2}(\eta-\eta_{0})$, which
describes the time evolution
of the sphaleron during its classical decay.

Of course,  these arguments do not prove that the sphaleron
relates to the absolute minimum of energy for static, nonvacuum
solutions. Notice however, that (\ref{8:1}) is the only
static, nonvacuum, $SO(4)$ symmetric solution \cite{henneaux}.
It is therefore very plausible that this solution does indeed minimize 
the energy.

Another handy form
for the sphaleron solution (\ref{8:1}) can be achieved as follows.
Introduce the
left and right invariant 1-forms on $S^{3}$, 
\be
\omega^{a}_{L}=\frac{i}{2}tr (\tau^{a}U^{(1)}dU^{(1)-1}),\ \ \ \
\omega^{a}_{R}=\frac{i}{2}tr (\tau^{a}U^{(1)-1}dU^{(1)}),
                                                       \label{10}
\ee
which satisfy the Maurer-Cartan equations
\be
d\omega^{a}+
\varepsilon_{abc}\ \omega^{b}\wedge\omega^{c}=0.        \label{11}
\ee
This allows us to represent the  field (\ref{8:1}) as
\be
A^{(sp)}=T_{a}\ \omega^{a}_{L}.              \label{12}
\ee

It is worth noting the following feature of this solution:
the sphaleron configuration consists of the gauge field alone.
This, together with the high symmetry of the solution,
will be of crucial importance for the analysis below.
It is well known that in the Minkowski space, the existence
of static, finite energy solutions for the pure gauge field
is ruled out by the scaling arguments. 
However, these arguments do not generally apply in curved spacetime,
since the invariance with respect to the rescaling of the 
coordinates, $\x\rightarrow\lambda\x$, is broken
by the  curvature. It is therefore gravity
which ensures the existence of the static sphaleron solution. 
Since the spacetime geometry is homogeneous and isotropic,
the sphaleron inherits the same
symmetries, such that, for instance, its
energy-momentum tensor has the manifest $SO(4)$-symmetric structure.

Let us mention also the following point:
since the Yang-Mills equations are conformally
invariant, and the geometry (\ref{1}) is conformally flat,
there exists the Minkowski space counterpart of the sphaleron solution.
However, this flat spacetime solution is, of course, no longer static.
Specifically, if one chooses
the conformal factor of the metric (\ref{1}) as
$\A=\A(\eta,\xi)=(\cos\eta +\cos\xi)^{-1}$ and introduces
the new coordinates $t\pm r=\tan((\eta\pm\xi)/2)$, the 
metric assumes the standard flat form \cite{cosmsphal}.
In the new coordinates,
the sphaleron field (\ref{8:1}) becomes a member of the family of the 
elliptic solutions \cite{alfaro}, which describe spherical
shells of the Yang-Mills radiation in the Minkowski space \cite{sing}.
In this sense, one can think of the sphaleron as being a
radiative solution which is rendered static by gravity.

We will also need  some knowledge about instantons
in the model. We first note that
the vacua and the sphaleron, since they are static, can be 
regarded as Euclidean solutions. Next, let us
pass to the imaginary time in Eq.(\ref{9}), 
$\eta\rightarrow -i \tau$,
\be
\left(\frac{dh}{d\tau}\right)^{2}-(h^{2}-1)^{2}=
-\varepsilon.                                     \label{inst}
\ee
Apart from the static solutions, this equation admits also 
the interpolating solutions for $\varepsilon=0$, 
such that $h(-\infty)=-1$, $h(\infty)=1$.
The periodic  solutions exist for 
$0<\varepsilon <1$,
and the corresponding  period is bounded from below,
\be
 \tau >\sqrt{2}\pi.                                 \label{inst:1}
\ee
Other known instanton solutions on $S^{3}$
can be obtained from the flat space BPST instantons by making use
of the conformal invariance of the YM equations \cite{baal}.

\newp

\section{The sphaleron transition rate}
\setcounter{equation}{0}

Consider the low energy excitations over the $k$-th topological
vacuum, 
$A^{(k)}(\x)\rightarrow
A^{(k)}(\x,t)= A^{(k)}(\x)+\delta A(\x,t)$. 
If the energy is  small  compared to
the barrier height,   $3\pi^{2}/g^{2}\A$, then the
excitations over the distinct vacua are classically independent. 
There is a nonzero amplitude for
the quantum tunneling between distinct sectors,
however the corresponding  probability is exponentially small.
On the perturbative level, 
one can consider the excitations in each sector 
independently.  The energy of the ground state
excitation in each sector
is $1/\A$ (see Eq.(\ref{4:18}) below).
The necessary condition for the smallness of the energy of
the excitations is
therefore $g^{2}/3\pi^{2}\ll 1 $. 

Consider the zeroth topological sector and assume a
thermal distribution for the states in this sector.
There is a finite probability for  such a thermal system to decay, 
both 
because of the underbarrier tunneling 
and due to the overbarrier thermal excitation.
According to the Langer-Affleck theory
of the metastable phase \cite{langer},
the decay rate is proportional
to the imaginary part of the free energy. To estimate the latter,
it is convenient to use the path integral approach
(the precise
definition of the path integration procedure will be given
in the next section).

The partition function of the gauge field is 
\be                                            \label{2:3}
Z=\exp(-\beta F)=
\int d[A] \ \exp(-S_{E}[A]).
\ee
In this expression, $S_{E}[A]$ is the Euclidean action of the
gauge field
in the static Riemannian space $(S^{1}\times S^{3},\g)$, with $\g$
being the analytic continuation of the metric (\ref{1}) to the
imaginary time,
$\eta\rightarrow -i\tau$, $\tau\in [0,\beta ]$.
In the weak coupling limit, one can approximate the partition
function by the sum over the classical extrema, $A^{\{ j\}}$, as
$Z\simeq\sum_{j}Z_{j}$,
where the semiclassical contribution of the $j$-th extremum is
\be                                         \label{2:4}
Z_{j}=\exp(-\beta F_{j})=\exp(-S[A^{\{ j\}}])
\int d[\varphi] \exp(-\delta^{2}S_{j}).    
\ee
The action for the 
fluctuations around the $j$-th extremum,
$A^{\{j\}}\rightarrow A^{\{j\}}+\varphi$, can be represented as
\be                                                \label{2:5}
S[A^{\{ j\}}+\varphi]=
S[A^{\{ j\}}]+\delta^{2}S_{j}+\ldots ,\ \ \ \ \
\delta^{2}S_{j}=\int_{0}^{\beta}d\tau\int_{S^{3}}
(\varphi,\hat{\D}_{j}\varphi) \ \sqrt{\g}d^{3}x,   
\ee
where $\hat{\D}_{j}=\hat{\D}[A^{\{ j\}}]$ is the Gaussian fluctuation
operator. This gives the one loop expression for the partition
function,
\be                                              \label{2:6}
Z\simeq\sum_{j}Z_{j}=\sum_{j}
\frac{\exp(-S[A^{\{ j\}}])}{\sqrt{Det(\hat{\D}_{j})}}.
\ee
Assume  that this sum is dominated by 
two terms, $Z\simeq Z_{0}+Z_{1}$, where $Z_{0}$ and $Z_{1}$ are
the contributions of the vacuum and the sphaleron, respectively.
Other periodic instantons that could
exist for a given value of $\beta$ are assumed to have a large
action.

The sphaleron fluctuation operator $\hat{\D}_{1}$ has at least one
negative eigenvalue $\omega^{2}_{-}<0$. Under the condition
specified by the lower bound in Eq.(\ref{2:11}) below, 
there is only one negative eigenvalue.  This implies that 
$Z_{1}$ is purely imaginary, and the free energy of the
whole system picks up the imaginary part
\be                                                  \label{2:8}
{\rm Im}(\beta F)=-{\rm Im}\ \ln Z\simeq
-{\rm Im} \ln\left(Z_{0}(1+\frac{Z_{1}}{Z_{0}})\right)\simeq
-\frac{1}{Z_{0}}{\rm Im}\ Z_{1}.                  
\ee

According to Langer \cite{langer},the imaginary part of the 
free energy is to be interpreted as giving rise to the decay rate of the
unstable phase built over the perturbative vacuum as
\be                                                      \label{2:9}
\Gamma=\frac{|\kappa|}{\pi T}\ {\rm Im}\ F,          
\ee 
where the damping constant $\kappa$ is the real time decay
rate of the sphaleron configuration in the heat bath.
In the weak coupling limit one has  
$|\kappa|=|\omega_{-}|$ \cite{arnold},
which finally determines the decay rate to be
\be                                                  \label{2:10}
\Gamma=-\frac{|\omega_{-}|}{\pi}\frac{{\rm Im}Z_{1}}{Z_{0}}.
\ee
This formula holds in the following  range of temperatures
\cite{langer}:
\be                                                                            
          \label{2:11}
\frac{|\omega_{-}|}{2\pi}<\frac{1}{\beta}\ll
\frac{3\pi^{2}}{g^{2}}.
\ee
The lower bound rules out the 
periodic instantons which play the leading role at low temperatures 
\cite{langer} (for the cosmological sphaleron
one has $|\omega_{-}|=\sqrt{2}$, 
such that this condition is opposite to that specified by
(\ref{inst:1})). 
The upper bound is the sphaleron energy which 
must exceed the temperature, otherwise the system would not be
metastable. Notice that this formula involves only
conformally invariant quantities, 
the conformal factor $\A$ drops out. 
Another crucial assumption 
is the weak coupling limit, $g^{2}/4\pi\ll 1$. First of all, this ensures
the very existence of the thermal ensemble. In addition, it justifies the
validity of the Gaussian approximation, and, moreover, leads to the
weak damping limit for $\kappa$.

\newp

\section{The path integration procedure}
\setcounter{equation}{0}

We outline below the main steps of the path integration procedure.
It is worth noting that the gauge field theory on $S^{3}$ resembles 
that on $S^{4}$. 
We shall therefore mainly follow the approach given in
Refs.\cite{polyakov}--\cite{osborn}.

Passing to the imaginary time, the spacetime
metric (\ref{1}) becomes
\be
ds^{2}=\A^{2}(d\tau^{2}+d\Omega^{2}_{3}),              \label{3:1}
\ee
where $\tau\in [0,\beta]$. We assume that the coordinates are
dimensionless, implying that $[\A]=[L]$.
Notice that $1/\beta$ is the {\it conformal} temperature. 
The physical temperature $T$ is defined with respect to the physical
time, $\A\tau$, such that $T=1/\beta\A$.

The Euclidean action of the gauge field is
\be
S_{E}[A]=\frac{1}{2g^{2}}tr\int F_{\mu\nu}F^{\mu\nu}
\ \sqrt{\g}d^{4}x > 0.                               \label{3:1:1}   
\ee
Consider small fluctuations around the $j$-th extremum of the action
$A_{\mu}^{\{j\}}\rightarrow A_{\mu}^{\{j\}}+\varphi_{\mu}$, 
where the values $j=1,0$ refer to the sphaleron and the vacuum
configurations,  respectively
(we shall omit this index where possible). 
The infinitesimal gauge transformations act as
$\varphi_{\mu}\rightarrow \varphi_{\mu}^{(\alpha)}=
\varphi_{\mu}+D_{\mu}\alpha$, 
where $D_{\mu}\alpha=\nabla_{\mu}\alpha-i[A_{\mu},\alpha]$, and
$\alpha$ is a Lie algebra valued scalar field.
Define the following operators:
\be                                                 \label{3:2:1}
\hat{\D}\varphi^{\nu}=\hat{\M}\varphi^{\nu}+D^{\nu}
(D_{\sigma}\varphi^{\sigma}),\ \ \ \
\hat{\M}\varphi^{\nu}=-D_{\sigma}D^{\sigma}\varphi^{\nu}
+R^{\nu}_{\sigma}\varphi^{\sigma}+
2i[F^{\nu}_{\ \sigma},\varphi^{\sigma}]. 
\ee
These are the vector fluctuation operator and the gauge fixed
fluctuation operator, respectively. Introduce also the
Faddeev-Popov operator
\be
\hat{\M}^{FP}\alpha=-D_{\sigma}D^{\sigma}\alpha.     \label{3:3}
\ee
In these formulas
$R^{\nu}_{\sigma}$ is the Ricci tensor for the geometry
(\ref{3:1}), and $F_{\nu\sigma}$ is the background gauge
field tensor. These operators are self-adjoint (symmetric)
with respect to the scalar products
\be
\langle\varphi,\varphi'\rangle =
2\ tr \int \varphi^{\sigma}\varphi_{\sigma}'\sqrt{\g}d^{4}x,\ \ \
\langle\alpha,\alpha'\rangle =
2\ tr \int \alpha\alpha'\sqrt{\g}d^{4}x.      \label{3:4}
\ee
The norms are
$||\varphi||=\sqrt{\langle\varphi,\varphi\rangle}$, and 
$||\alpha||=\sqrt{\langle\alpha,\alpha\rangle}$.
The action can be expanded as
\be
S_{E}[A+\varphi]\approx S_{E}[A]+
\frac{1}{2g^{2}}
\langle\varphi,\hat{\D}\varphi\rangle.       \label{3:5}
\ee
The Gaussian path integral is
\be
Z=\exp (-S_{E}[A])\int d^{FP}[\varphi]\exp\left(-
\frac{1}{2g^{2}}
\langle\varphi,\hat{\D}\varphi\rangle\right),       \label{3:6}
\ee
where the Faddeev-Popov measure is
\be                                                 \label{3:7}
d^{FP}[\varphi]=d[\varphi]{\cal G}(\varphi)
{\cal F}(\varphi),\ \ \ \ \ \
1={\cal G}(\varphi)\int d[\alpha]
{\cal F}(\varphi^{(\alpha)}),                                        
\ee
and ${\cal F}$ is the gauge fixing function.
The fluctuations $\varphi$ can be decomposed as
\be                                                                                                  \label{3:7:1}
\varphi_{\mu}=D_{\mu}\alpha+\xi_{\mu},
\ee
where the pure gauge part $D_{\mu}\alpha$ is annihilated
by $\hat{\D}$, and
$\xi_{\mu}$ is orthogonal to all gauge modes,
$D_{\sigma}\xi^{\sigma}=0$. The fields $\xi_{\mu}$ and
$\alpha$ can be expanded with respect to the eigenfunctions
of $\hat{\D}$ and $\hat{\M}^{FP}$:
\be                                                                                                           \label{3:10}
\xi^{\mu}=\sum_{k} C_{k}\xi_{k}^{\mu},\ \ \ 
\alpha=\sum_{n} B_{n}\alpha_{n};\ \ \ \ \ \ \ \ \ \ \
\hat{\D}\xi^{\nu}_{k}=\lambda_{k}\xi^{\mu}_{k},\ \ \
\hat{\M}^{FP}\alpha_{n}=q_{n}\alpha_{n}. 
\ee
The gauge fixing function is chosen to be 
\be                                          \label{3:16}
{\cal F(\varphi)}=\exp\left\{-\frac{1}{2g^{2}}
\langle D_{\sigma}\varphi^{\sigma},
D_{\mu}\varphi^{\mu}\rangle\right\}
=\exp\left\{
-\frac{1}{2g^{2}}
\sum_{n}B_{n}^{2}q_{n}^{2}||\alpha_{n}||^{2}\right\},
\ee
and the integration measure is the square root of the determinant
of the metric on the function space \cite{osborn}:
\be                                                \label{3:16:a}
d[\varphi]=\prod_{k}\frac{\mm}{\sqrt{2\pi}g}dC_{k}||\xi^{\mu}_{k}||
\prod_{n}'\frac{\mm}{\sqrt{2\pi}g}dB_{n}\sqrt{q_{n}}||\alpha_{n}||,
\ \ \ \ \ \ \
d[\alpha]=\prod_{n}\frac{\mm^{2}}
{\sqrt{2\pi}g}dB_{n}||\alpha_{n}||.                                                \label{3:15}
\ee
Here $\mm/(\sqrt{2\pi}g)$ is a normalization factor with $\mm$
being an arbitrary normalization scale;
the prime indicates that terms with $q_{n}=0$ should be omitted.
Taking (\ref{3:7})-(\ref{3:16:a}) into account, 
the Gaussian path integral in (\ref{3:6}) reduces to 
\be
{\cal G}\int
\prod_{k}\frac{\mm dC_{k}}{\sqrt{2\pi}g}||\xi^{\mu}_{k}||
\prod_{n}'\frac{\mm dB_{n}}{\sqrt{2\pi}g}||\alpha_{n}||
\exp\left\{-\frac{1}{2g^{2}}\left(
\sum_{k}\lambda_{k}C_{k}^{2}||\xi^{\mu}_{k}||^{2}
+\sum_{n}q_{n}B_{n}^{2}||\alpha_{n}||^{2}\right)
\right\},                                            \label{3:19}
\ee
where $\sqrt{q_{n}}$ has been absorbed in $B_{n}$, and
\be                                                  \label{3:19:1}
{\cal G}^{-1}=\int \prod_{n}\frac{\mm^{2}}
{\sqrt{2\pi}g}dB_{n}||\alpha_{n}||
\exp\left\{
-\frac{1}{2g^{2}}
\sum_{n}B_{n}^{2}q_{n}^{2}||\alpha_{n}||^{2}\right\}.
\ee

Let us first apply these formulae to the sphaleron.
In this case, as we will see in the next section, the vector
fluctuation 
operator has one negative eigenvalue, $\lambda_{-}<0$, whereas
all the other eigenvalues $\lambda_{k}$, $q_{n}$ in (\ref{3:19}),
(\ref{3:19:1}) are positive (zero modes are absent). 
The integral over $C_{-}$ in (\ref{3:19}) 
can be defined by analytic continuation
\cite{callan}, and the result is
$\mm/2i\sqrt{|\lambda_{-}|}$. The rest are well-defined
Gaussian integrals. 
It is convenient to introduce the conformally invariant
dimensionless
operators
$\hat{M}$ and $\hat{M}^{FP}$ whose eigenvalues are
$\omega^{2}_{k}$ and $\omega^{2}_{n}$:
\be                                                    \label{4:0}
\hat{\M}=\frac{1}{\A^{2}}\hat{M},\ \ \
\hat{\M}^{FP}=\frac{1}{\A^{2}}\hat{M}^{FP};\ \ \ \ \ \ \ \ \
\lambda_{k}=\frac{\omega^{2}_{k}}{\A^{2}},\ \ \ 
q_{n}=\frac{\omega^{2}_{n}}{\A^{2}}. 
\ee
This implies that the partition function for the Gaussian
fluctuations around the sphaleron is 
\be
Z_{1}=\exp (-S_{E}[A^{(sp)}])
\frac{\mm\A}{2i\sqrt{|\omega_{-}|}}
\frac{Det(\hat{M}^{FP}_{1}/\mm^{2}\A^{2})}
{\sqrt{Det'(\hat{M}_{1}/\mm^{2}\A^{2}})},       \label{3:20}
\ee
where $Det'$ has all nonpositive eigenvalues omitted, and the
index 1 referring to the sphaleron is restored.
Notice that $Det'(\hat{M})$ must be computed
on the space of all vector fluctuations $\varphi^{\nu}$,
and not only for those
satisfying $D_{\sigma}\varphi^{\sigma}=0$.

Consider now the vacuum case.  As we will see below,
the ghost operator $\hat{M}^{FP}$ in this case has
three zero modes  $\alpha_{p}=\tau^{p}/2$ ($p=1,2,3$)
related to the global gauge rotations of the vacuum $A=0$. 
The norm of these modes is
\be                                          \label{3:20:1}
||\alpha_{p}||^{2}=\int_{0}^{\beta}d\tau\int_{S^{3}}
\sqrt{\g}d^{3}x=2\pi^{2}\beta\A^{4}.
\ee
The quantity ${\cal G}$ specified by (\ref{3:19:1}) then reads
\be                                             \label{3:17}
{\cal G}=\frac{1}{\Upsilon}\prod_{n}'\frac{q_{n}}{\mm^{2}}=
\frac{1}{\Upsilon}Det'\left(\frac{\hat{\M}^{FP}}{\mm^{2}}\right),
\ee
with
\be                                                 \label{3:18}
\Upsilon=
\prod_{p=1}^{ 3}\int
\frac{\mm^{2}}{\sqrt{2\pi}g}dB_{p}||\alpha_{p}||=
V_{SU(2)}\frac{\pi\sqrt{\pi}}{g^{3}}\mm^{6}\A^{6}\beta^{3/2},
\ee
where the integration over $B_{p}$ gives
the volume of the stability group, 
in our normalization it is $V_{SU(2)}=16\pi^{2}$.
In addition, there are three constant vector modes
annihilated by the vector
fluctuation operator:
$\xi_{p}^{\mu}=\delta_{0}^{\mu}\tau^{p}/2$,
$||\xi_{p}^{\mu}||^{2}=2\pi^{2}\beta\A^{2}$.
Under the gauge transformation generated by
\be                                                   \label{3:31}
U_{p}(\tau)=\exp\left(i\tau\frac{2\pi}{\beta}l\tau^{p}\right),
\ \ \ U_{p}(0)=U_{p}(\beta),\ \ \ \ 
\ l\in{\cal Z},\
\ee
these modes change according to
\be                                                   \label{3:32}
U_{p}(\tau):
\xi_{p}^{\mu}=\delta_{0}^{\mu}\frac{\tau^{p}}{2}\rightarrow
\delta_{0}^{\mu}\frac{\tau^{p}}{2}\left(1+
\frac{4\pi l}{\beta}\right).
\ee
This shows that the range of integration over
$dC_{p}$ in (\ref{3:19})
is finite, $C_{p}\in [0,4\pi/\beta]$. The contribution
of these three modes to (\ref{3:19}) therefore is
\be                                               \label{3:33}
\int\prod_{p=1}^{3}\frac{\mm dC_{p}}{\sqrt{2\pi}g}||\xi^{\mu}_{p}||=
\left(\frac{4\pi}{\beta}\right)^{3}
\frac{\pi\sqrt{\pi}}{g^{3}}\mm^{3}\A^{3}\beta^{3/2}.
\ee
All other eigenvalues are positive, which finally gives
for the fluctuations around the vacuum
\be                                               \label{3:34}
Z_{0}=\frac{4\pi}{\mm^{3}\A^{3}\beta^{3}}
\frac{Det'(\hat{M}^{FP}_{0}/\mm^{2}\A^{2})}
{\sqrt{Det'(\hat{M}_{0}/\mm^{2}\A^{2}})}.       
\ee
We shall omit below the factor $\mu_{0}$,
$\mu_{0}\A\rightarrow\A$, 
such that $\A$ will be understood as the radius of the universe
expressed in units of an {\sl arbitrary} length scale.

\newp

\section{Spectra of the fluctuation operators}
\setcounter{equation}{0}

To analyze the spectra of the conformally invariant operators 
$\hat{M}$ and $\hat{M}^{FP}$ defined by Eqs.(\ref{3:2:1}), 
(\ref{3:3}), (\ref{4:0}), one can put $\A=1$ in the line element.
We introduce the 1-form basis $\{\omega^{0},\omega^{a}\}$ on the
spacetime manifold,
where $\omega^{0}=d\tau$, and $\omega^{a}=\omega^{a}_{L}$ are the
left invariant 1-forms given by Eq.(\ref{10}). The metric is
\be
ds^{2}=\omega^{0}\otimes\omega^{0}+
\omega^{a}\otimes\omega^{a}.                         \label{4:1}
\ee
Let $\{e_{0},e_{a}\}$ be the corresponding dual tetrad;
here $e_{a}=e^{L}_{a}$ are the left invariant vector fields on
$S^{3}$. 
Introduce the right invariant fields
$e^{R}_{a}$ dual to the 1-forms $\omega^{a}_{R}$. 
The following commutation relations hold
\be
[e^{L}_{a},e^{L}_{b}]=2\varepsilon_{abc}e^{L}_{c},\ \ \ \
[e^{R}_{a},e^{R}_{b}]=2\varepsilon_{abc}e^{R}_{c},\ \ \ \
[e^{L}_{a},e^{R}_{b}]=0.\ \ \ \                       \label{4:2}
\ee
Let $\nabla_{0}$ and  $\nabla_{a}$ denote the covariant
derivatives along the tetrad vectors $\{e_{0},e_{a}\}$. 
The following tetrad rotation coefficients do not vanish:
\be
\nabla_{a}e_{b}=\varepsilon_{abc}e_{c},\ \ \ \ \ \ \
\nabla_{a}\omega^{b}=\varepsilon_{abc}\omega_{c}.    \label{4:3}
\ee
Let us represent the gauge field of the vacuum and the sphaleron as
\be                                               \label{4:4}
A^{\{j\}}=j\ \omega^{s}T_{s},
\ee
where $j=0,1$, respectively.
All this suggests to expand the fluctuations as \cite{Hosotani}
\be
\varphi=\phi^{0}\omega^{0}+\phi^{a}\omega^{a},
                                                  \label{4:5}
\ee
where $\phi^{0}$ and $\phi^{a}$ are the scalar and the
vector fluctuations, respectively.

Using Eqs.(\ref{3:2:1}), (\ref{3:3}), (\ref{4:1})--(\ref{4:5}),   
a straightforward calculation shows that the fluctuation
operators $\hat{M}_{j}$ $(j=0,1)$ decompose into the direct sum
of the
two operators acting on the scalar and the
vector fluctuations, respectively:
\be
\hat{M}_{j}\phi=
\left(\hat{M}_{j}^{scal}\phi^{0}\right)\omega^{0}
+\left(\hat{M}_{j}^{vec}\phi^{a}\right)\omega^{a}.
                                                         \label{4:6}
\ee
Here the scalar fluctuation operator $\hat{M}^{scal}$ 
formally coincides with the
ghost operator $\hat{M}^{FP}$ introduced in the previous section,
\be
\hat{M}_{j}^{scal}\phi^{0}=-
(\nabla_{0}\nabla_{0}+\nabla_{a}\nabla_{a})\phi^{0}-
j\left\{2\nabla_{s}[T_{s},\phi^{0}]+
[T_{s},[T_{s},\phi^{0}]]\right\}.                      \label{4:7}
\ee
The vector operator reads
$$
\hat{M}_{j}^{vec}\phi^{a}=-
(\nabla_{0}\nabla_{0}+
\nabla_{c}\nabla_{c}-4)\phi^{a}-
2\varepsilon_{abc}\nabla_{b}\phi^{c}+
$$
\be
+j\left\{
-2\nabla_{s}[T_{s},\phi^{a}]+
4\varepsilon_{abs}[T_{s},\phi^{b}]-
[T_{s},[T_{s},\phi^{a}]]\right\}.                      \label{4:8}
\ee
Each of these operators decomposes into the direct  sum of
a temporal and a spatial part, 
\be                                                    \label{4:8:1}              
\hat{M}=-\frac{\partial^{2}}{\partial\tau^{2}}+\hat{\cal M},
\ee
such that the problem reduces to the study of the corresponding
spatial
operators $\hat{{\cal M}}_{j}^{scal}$ and $\hat{{\cal M}}_{j}^{vec}$.

We now introduce
\be                                                                           \label{4:9}
\La_{a}=\frac{i}{2}e_{a}^{L},\ \
\tilde{\La}_{a}=\frac{i}{2}e_{a}^{R},
\ \ \ \ \vec{\La}^{2}=\La_{a}\La_{a}=\tilde{\La}_{a}\tilde{\La}_{a},
\ee
which are the $SO(4)$ angular momentum operators, since 
\be                                                                             \label{4:10}
[\La_{a},\La_{b}]=i\varepsilon_{abc}\La_{c},\ \ \ \ 
[\tilde{\La}_{a},\tilde{\La}_{b}]=i\varepsilon_{abc}\tilde{\La}_{c},
\ \ \ \ [\La_{a},\tilde{\La}_{b}]=0. 
\ee
The commuting operators are $\vec{\La}^{2}$, $\La_{3}$ and 
$ \tilde{\La}_{3}$. The  eigenvalues are similar to those for the
$SO(3)$ case,
but the angular momentum  can now assume both integer and 
half-integer values:
\be
\vec{\La}^{2}=l(l+1);\ \ \ l=0,\frac{1}{2},1,\frac{3}{2},\ldots ;
\ \ \ \La_{3}=m,\  \tilde{\La}_{3}=\tilde{m};\ \  m,
\tilde{m}=-l,-l+1,\ldots l.             \label{4:11}
\ee
Next, expanding the fluctuations over the basis of the  Lie algebra, 
$\phi^{0}=\phi^{0}_{p}T_{p}$, $\phi^{a}=\phi^{a}_{p}T_{p}$,
we define the spin and isospin operators $\vec{\Sa}$ and $\vec{\Ta}$ by
\be                                                  \label{4:12}
\Sa_{a}\phi^{b}_{p}=\frac{1}{i}\varepsilon_{abc}\phi^{c}_{p},\ \ \ 
\Ta_{p}\phi^{a}_{r}=\frac{1}{i}\varepsilon_{prs}\phi^{a}_{s},\ \ \ 
\Ta_{p}\phi^{0}_{r}=\frac{1}{i}\varepsilon_{prs}\phi^{0}_{s},
\ee
which satisfy the usual commutation relations. One has  
$\vec{\Sa}^{2}=\vec{\Ta}^{2}=2$, which corresponds to the 
 unit spin and unit isospin, respectively. 

Using Eqs.(\ref{4:7})-(\ref{4:12}) one can represent the spatial
operators for the 
fluctuations around the sphaleron $(j=1)$ and the vacuum ($j=0$) 
configurations as 
$$
\hat{{\cal M}}_{1}^{vec}=2\left(\vec{\La}^{2}+
(\vec{\La}+\vec{\Sa}+\vec{\Ta})^{2}-1\right),\ \ \ \ \ \ \ \ \
\hat{{\cal M}}_{1}^{scal}=2\left(\vec{\La}^{2}+
(\vec{\La}+\vec{\Ta})^{2}-1\right),
$$
\be
\hat{{\cal M}}_{0}^{vec}=2\left(\vec{\La}^{2}+
(\vec{\La}+\vec{\Sa})^{2}\right),\ \ \ \ \ \ \ \ \
\hat{{\cal M}}_{0}^{scal}=4\vec{\La}^{2}.                \label{4:13}
\ee
The fluctuation operators therefore reduce to the combinations
of the angular momentum operators whose spectra can
be analyzed by the usual methods. Let us illustrate the procedure
for the operator $\hat{\cal M}_{1}^{vec}$.

We represent the operator in the form
$\hat{{\cal M}}_{1}^{vec}=2\left(\vec{\La}^{2}+
\vec{\Ja}^{2}-1\right)$,
where $\vec{\Ja}=\vec{\La}+\vec{\Ka}$, and
$\vec{\Ka}=\vec{\La}+\vec{\Ta}$.
The commuting operators are 
$\vec{\La}^{2}$, $\vec{\Ja}^{2}$, $\tilde{\La}_{3}$ and $\Ja_{3}$,
such that the eigenvalues of
$\hat{\cal M}_{1}^{vec}$ and their degeneracies
read $\omega^{2}=2\{l(l+1)+j(j+1)-1\}$ and $d=\nu (2l+1)(2j+1)$,
respectively.
Here $\nu$ is the degeneracy factor associated with
the several possibilities to obtain a given value of $j$
for a given value of $l$.

To find $\nu$, let us fix a value $l\geq 2$.
$\vec{\Ka}$ is the sum of the two unit
angular momenta, such that $K=0,1,2$.
For $K=2$ the possible values of $j$ are
$l-2,l-1\ldots, l+2$, for $K=1$ one obtains
$j=l-1, l,l+1 $ and for $K=0$ one has $j=l$.
The values $j=l\pm2$ therefore appear only when $K=2$,
the values $j=l\pm1$ are possible when $K=1$ or $K=2$,
and there are three
different ways to get $j=l$, that is, when $K=0,1,2$.
Thus, if we write $j=l+\sigma$, where $\sigma=0,\pm 1,\pm 2$, 
the degeneracies $\nu$ are as follows:
$\nu=3$ for $\sigma=0$, $\nu=2$ for $\sigma=\pm1$ and
$\nu=1$ for $\sigma=\pm 2$.
As a result, the eigenvalues of $\hat{\cal M}_{1}^{vec}$
and their degeneracies can be represented as
\be
\omega^{2}=(2l+\sigma+1)^{2}+\sigma^{2}-3,\ \ \
d=(3-|\sigma|)(2l+1)(2l+2\sigma+1),\ \ \
l\geq 2,\ \ \sigma=0,\pm1,\pm2.         \label{4:14}
\ee

For $l<2$, the factor $\nu$ changes.
For instance for $l=0$ one has $j=K$, such that
$\nu=1$ for $\sigma=2,1,0$ and $\nu=0$ for $\sigma=-1,-2$.
Similarly for $l=1/2$ one obtains $\nu=\{1,2,2,0,0\}$
for $\sigma=\{2,1,0,-1,-2\}$, respectively.
For $l=1$ one has $\nu=\{1,2,3,1,0\}$ and for $l=3/2$
the result is $\nu=\{1,2,3,2,0\}$.

Introducing the new quantum number,
$n=2l+\sigma+1$, 
the whole spectrum can be given in the following compact form:
\be
\hat{\cal M}_{1}^{vec}:\ \ \ \
\omega^{2}=n^{2}+\sigma^{2}-3,\ \ \ \ \ \
d=\nu(n^{2}-\sigma^{2}),\ \ \
\ \ \ \ \ \sigma=0,\pm 1,\pm 2;                   \label{4:15}
\ee
where $\nu=3-|\sigma|$ for $n\geq 3$, 
and $\nu=n\delta_{0\sigma}+\delta_{n2}\delta_{1|\sigma|}$
for $n=1,2$.
For the other fluctuation operators one similarly obtains
$$
 \hat{\cal M}_{1}^{scal}:\ \
 \omega^{2}=n^{2}+\sigma^{2}-3, \ \ \ \ \
 d=n^{2}-\sigma^{2}, \ \
 \sigma = 0,\pm 1,\ \ \\
 n\geq 2;
$$
$$
\hat{\cal M}_{0}^{vec}: \ \
\omega^{2}=n^{2}+\sigma^{2}-1,\ \ \ \ \
d=3(n^{2}-\sigma^{2}),\ \
\sigma=0,\pm 1,\ \
n\geq 2;
$$
\be
\hat{\cal M}_{0}^{scal}:\ \
\omega^{2}=n^{2}-1,\ \ \ \ \
d=3n^{2},\ \  n\geq 1.                    \label{4:16}
\ee

All these eigenvalues are positive for $n\geq 2$. 
When $n=1$, the operator
$\hat{\cal M}_{1}^{vec}$ has one negative mode with 
eigenvalue $\omega_{-}^{2}=-2$.
It is worth noting 
that the sphaleron does not possess
any zero modes. This
can be understood as follows: 
the sphaleron solution completely shares the
symmetries of the 3-space -- both are $SO(4)$
symmetric. The sphaleron is therefore invariant under 
spatial translations and rotations. In this case, 
all zero modes must be of pure
gauge origin, but the gauge is completely fixed.

For $n=1$, the vacuum operator $\hat{\cal M}_{0}^{scal}$
has three zero modes. As is seen from (\ref{4:13}), their
eigenfunctions are just constants
discussed in the previous section: $\xi^{\mu}_{p}=\delta^{\mu}_{p}
\tau^{p}/2$. Since the spectra of
$\hat{{\cal M}}^{scal}$ and $\hat{{\cal M}}^{FP}$ 
coincide (but these operators act in the different spaces),
$\hat{{\cal M}}_{0}^{FP}$
also has three constant zero modes, $\alpha_{p}=\tau^{p}/2$, which
have been discussed above.

The next step is to pick up the physical modes. The partition
functions $Z_{j}$ specified by (\ref{3:20}) and (\ref{3:34})
involve the ratios of the determinants,
such that some of the eigenvalues cancel.
Notice that (\ref{4:6}) implies that
$Det'(\hat{M})=Det'(\hat{M}^{scal})Det'(\hat{M}^{vec})$,
whereas $Det'(\hat{M}^{scal})=Det'(\hat{M}^{FP})$. 
The partition functions therefore are \cite{polyakov}
\be                                                \label{4:16a}
Z_{1}=
\exp (-S_{E}[A^{(sp)}])
\frac{\A}{2i\sqrt{|\omega_{-}|}}
\sqrt{\frac{Det'(\hat{M}_{1}^{scal}/\A^{2})}
{Det'(\hat{M}_{1}^{vec}/\A^{2})}},\ \ \
Z_{0}=
\frac{4\pi}{\A^{3}\beta^{3}}
\sqrt{\frac{Det'(\hat{M}_{0}^{scal}/\A^{2})}
{Det'(\hat{M}_{0}^{vec}/\A^{2})}}.
\ee
Thus the physical oscillator modes constitute the part of the
spectrum of $\hat{M}^{vec}$ that remains after the subtraction
of all eigenvalues of $\hat{M}^{scal}$. Since all eigenvalues
of $\hat{M}^{scal}$ are contained in the spectrum of $\hat{M}^{vec}$
(besides the zero modes of $\hat{M}^{vec}_{0}$),
such a subtraction results in the change of the degeneracy
factors of the eigenmodes of $\hat{M}^{vec}$. 
Using (\ref{4:15}), (\ref{4:16}), one obtains
the eigenvalues and the degeneracies, $\{\omega^{2},d\}$,
of the physical oscillators:
\be
\hat{\cal M}_{1}^{vec}/\hat{\cal M}_{1}^{scal}:\
\{-2,1\},\ \{1,4\},\ 
\{n^{2}+\sigma^{2}-3,2(n^{2}-\sigma^{2})\}, \ \ \
n\geq 3,\ \sigma=0,1,2;                            \label{4:17}
\ee
also 
\be
\hat{\cal M}_{0}^{vec}/\hat{\cal M}_{0}^{scal}:\
\{0,-3\},\ \{n^{2},6(n^{2}-1)\}, \ \ \ n\geq 2.      \label{4:18}
\ee
Notice that $\sigma$ in (\ref{4:17}) assumes three
values which correspond to the three ``colours'' of the
``gluon'', and the coefficient 2 in the degeneracy
factor refers to its two polarizations, whereas in (\ref{4:18})
all (2 spin)$\times$(3 isospin) gluon degrees of freedom are 
absorbed by the coefficient 6 in the degeneracy factor.
Taking into account the temporal part of the fluctuation operators,
$-\frac{\partial^{2}}{\partial\tau^{2}}$,
whose eigenvalues are the Matsubara frequencies,
$4\pi^{2}l^{2}/\beta^{2}$,
we obtain the following contribution of each physical
oscillator, $\{\omega^{2},d\}$, into the partition function:
\be                                                  \label{4:19}
\prod_{l=-\infty}^{\infty}
\left\{\frac{1}{\A^{2}}
\left(\frac{4\pi^{2}l^{2}}{\beta^{2}}+\omega^{2}\right)
\right\}^{-d/2}.
\ee
For $\omega^{2}<0$ or $\omega^{2}=0$ the term with $l=0$ in this
product should be omitted, since the corresponding negative or
zero modes have already been taken into account. 
Notice that the negative degeneracy for the zero mode in (\ref{4:18})
arises simply because this mode is  not in the spectrum of
$\hat{\cal M}^{vec}_{0}$ and only in that
of $\hat{\cal M}^{scal}_{0}$.

We are now in a position to give the formal
closed expressions for the
partition functions. 
For the fluctuations around the sphaleron we obtain
$$
\frac{\exp(-S_{E}[A^{(sp)}]) }{Z_{1}}=
i2\frac{\sqrt{2}}{\A}\prod_{l=1}^{\infty}
\left\{\frac{1}{\A^{2}}
\left(\frac{4\pi^{2}l^{2}}{\beta^{2}}-2\right)\right\}\times
\prod_{l=-\infty}^{\infty}
\left\{\frac{1}{\A^{2}}
\left(\frac{4\pi^{2}l^{2}}{\beta^{2}}+1\right)\right\}^{2}\times
$$
\be
\times\prod_{l=-\infty}^{\infty}
\prod_{\sigma=0,1,2}\prod_{n=3}^{\infty}
\left\{\frac{1}{\A^{2}}
\left(\frac{4\pi^{2}l^{2}}{\beta^{2}}+n^{2}+\sigma^{2}-3\right)
\right\}^{n^{2}-\sigma^{2}},                        \label{4:20}
\ee
and for the vacuum
\be
\frac{1}{Z_{0}}=\frac{\A^{3}\beta^{3}}{4\pi}
\prod_{l=1}^{\infty}
\left\{\frac{1}{\A^{2}}
\left(\frac{4\pi^{2}l^{2}}{\beta^{2}}\right)\right\}^{-3}\times
\prod_{l=-\infty}^{\infty}
\prod_{n=2}^{\infty}
\left\{\frac{1}{\A^{2}}
\left(\frac{4\pi^{2}l^{2}}{\beta^{2}}+n^{2}\right)
\right\}^{3(n^{2}-1)}.                              \label{4:21}
\ee

\newp

\section{Evaluation of determinants and the transition rate}
\setcounter{equation}{0}

We use the zeta function techniques
(see \cite{zeta}-\cite{abram}, and references therein)
to regularize and evaluate the infinite
products entering Eqs.(\ref{4:20}), (\ref{4:21}).
The key steps of our analysis are presented in this
section, whereas a large number of the technical details are
given in the Appendix.

The basic zeta function relation reads
\be                                                      \label{5:1}
\prod_{n}\left(\frac{\lambda_{n}}{\mu}\right)^{d_{n}}=
\exp \{-\zeta'(0)-\ln\mu\ \zeta(0)\},                                          
\ee
where
\be                                                     \label{5:2}
\zeta(s)=\sum_{n}\frac{d_{n}}{(\lambda_{n})^{s}}=
\frac{1}{\Gamma(s)}\int_{0}^{\infty}t^{s-1}\sum_{n}d_{n}
\exp(-t\lambda_{n})dt,                                                         
\ee
which in fact should be regarded as the definition of the product.

We start by applying this to the
zero mode contribution  in (\ref{4:21}):
\be                                                     \label{5:3}
\prod_{l=1}^{\infty}
\left\{\frac{1}{\A^{2}}
\left(\frac{4\pi^{2}l^{2}}{\beta^{2}}\right)\right\}^{-3}=
\prod_{l=1}^{\infty}
\left(\frac{2\pi l}{\A\beta}\right)^{-6}.
\ee
The scale factor $\mu$ is given by $\mu=\A\beta/2\pi$,
and the zeta function is
\be                                                    \label{5:4}
\zeta_{0}(s)=-6\sum_{l=1}^{\infty}\frac{1}{l^{s}}=-6\zeta_{R}(s),
\ \ \Rightarrow\ \ \
\zeta_{0}(0)=3,\ \ \ \zeta_{0}'(0)=3\ln 2\pi,
\ee
where $\zeta_{R}(s)$ is the Riemann zeta function.
This gives
\be                                                   \label{5:5}
\prod_{l=1}^{\infty}
\left\{\frac{1}{\A^{2}}
\left(\frac{4\pi^{2}l^{2}}{\beta^{2}}\right)\right\}^{-3}=
\frac{1}{\A^{3}\beta^{3}},
\ee
such that the overall
contribution of the zero modes together with their Matsubara
excitations into (\ref{4:21}) is
\be                                                 \label{5:5:1}
\frac{\A^{3}\beta^{3}}{4\pi}\prod_{l=1}^{\infty}
\left\{\frac{1}{\A^{2}}
\left(\frac{4\pi^{2}l^{2}}{\beta^{2}}\right)\right\}^{-3}=
\frac{1}{4\pi}.
\ee

Next, consider the product in (\ref{4:20}) due to the negative mode.
The corresponding zeta function is
\be                                                   \label{5:5:2}
\zeta_{-}(s)=\sum_{l=1}^{\infty}\left(l^{2}-
\frac{\beta^{2}}{2\pi^{2}}\right)^{-s},\ \Rightarrow\ \
\zeta_{-}(0)=-\frac{1}{2},\ \ \ \zeta_{-}'(0)=-\ln\left(
\frac{2\sqrt{2}\pi}{\beta}\sin(\frac{\beta}{\sqrt{2}})\right)
\ee
(see Appendix, Eq.(\ref{a8})). The normalization factor is
$\mu=(\beta\A/2\pi)^{2}$, which yields
\be
2\frac{\sqrt{2}}{\A}
\prod_{l=1}^{\infty}
\left\{\frac{1}{\A^{2}}
\left(\frac{4\pi^{2}l^{2}}{\beta^{2}}-2\right)\right\}=
4\sin\frac{\beta}{\sqrt{2}}.                 \label{5:8}
\ee

Now we want to take into account the contribution of the positive
field modes. We introduce
the spatial zeta function associated with the positive physical
oscillators (\ref{4:17}), (\ref{4:18})
\be                                                 \label{5:18}
\zeta_{spat}(s)=
2+\sum_{\sigma=0,1,2}\ \sum_{n=3}^{\infty}
\frac{n^{2}-\sigma^{2}}{(n^{2}+\sigma^{2}-3)^{s}}-
3\sum_{n=2}^{\infty}\frac{n^{2}-1}{(n^{2})^{s}},
\ee
such that
\be                                                  \label{5:18:1}
\sqrt{
\frac{Det'\hat{\cal M}_{1}^{scal}}{Det\hat{\cal M}_{1}^{vec}}
\frac{Det\hat{\cal M}_{0}^{vec}}{Det'\hat{\cal M}_{0}^{scal}}
}=\exp\{\zeta_{spat}'(0)\}.
\ee
One has
\be                                                 \label{5:18a}
\zeta_{spat}(s)=
\frac{1}{\Gamma(s)}
\int_{0}^{\infty}dt\ t^{s-1}\Theta_{spat}(t),              
\ee
where the heat kernel is
\be
\Theta_{spat}(t)=2\exp\{-t\}+\sum_{\sigma=0,1,2}\sum_{n=3}^{\infty}
(n^{2}-\sigma^{2})\exp\{-t(n^{2}+\sigma^{2}-3)\}-
3\sum_{n=2}^{\infty}
(n^{2}-1)\exp\{-t n^{2}\}.                               \label{5:11}
\ee
We shall need the asymptotic expansion of this function for small $t$
\be
\Theta_{spat}(t)\sim\frac{1}{(4\pi t)^{3/2}}
\sum_{r=0,1/2,1,\ldots}C_{r}t^{r}.                      \label{5:12}
\ee
The computation of the coefficients $C_{r}$ for $r\leq 2$ is
performed in the Appendix:
\be
C_{0}=C_{1/2}=C_{1}=0,\ \ C_{3/2}=-2(4\pi)^{3/2},\ \
C_{2}=22\pi^{2}.                             \label{5:14}
\ee
Next, we introduce the thermal zeta function  related to
the spatial zeta function
\be
\zeta_{\beta}(s)=
\frac{1}{\Gamma(s)}\int_{0}^{\infty}t^{s-1}\sum_{l=-\infty}^{\infty}
\exp\left\{-\left(\frac{2\pi l}{\beta}\right)^{2}t\right\}
\Theta_{spat}(t)dt.                                     \label{5:10}
\ee
As a result, we can collect all parts
together and represent the expression for the sphaleron transition
rate in the following form:
\be                                                    \label{5:9}
\Gamma=-\frac{\omega_{-}}{\pi}\frac{{\rm Im} Z_{1}}{Z_{0}}=
\frac{1}{8\sqrt{2}\pi^{2}\sin(\beta/\sqrt{2})}
\exp\left\{-\frac{3\pi^{2}}{g^{2}}\beta+
\zeta_{\beta}'(0)+\ln\A^{2}\ \zeta_{\beta}(0)\right\}.                   
\ee
Here the prefactor on the right hand side includes the 
contribution of the zero modes and the negative mode.
The exponent contains
the one-loop contribution of the positive oscillator modes, and
the classical sphaleron action
\be                                                      \label{5:9a}
S_{E}[A^{(sp)}]=\frac{3\pi^{2}}{g^{2}}\beta.
\ee

To compute the quantities $\zeta_{\beta}'(0)$ and $\zeta_{\beta}(0)$
entering Eq.(\ref{5:9}), we represent the 
heat kernel  in (\ref{5:11}) symbolically as
\be                                                   \label{5:15}
\Theta_{spat}(t)=\sum_{\omega}\exp\{-\omega^{2}t\}, \ \ \ \ \
\omega^{2}>0.
\ee
Then the values of $\zeta_{\beta}(0)$ and
$\zeta_{\beta}'(0)$ are given by (see Appendix)
\be                                                    \label{5:20}
\zeta_{\beta}(0)=\frac{C_{2}\beta}{16\pi^{2}},\ \ \ \ \
\zeta_{\beta}'(0)=
\left\{\frac{(2-2\ln 2) C_{2}}{16\pi^{2}}-
{\rm PP}\zeta_{spat}(-\frac{1}{2})\right\}\beta-
2\sum_{\omega}\ln\left(1-e^{-\beta \omega}\right),      
\ee
where the coefficient $C_{2}$ is specified by (\ref{5:14}).

All this allows us to represent the transition rate (\ref{5:9}) as
\be                                                   \label{5:20a}
\Gamma=
\frac{1}{8\sqrt{2}\pi^{2}\sin(\beta/\sqrt{2})}
\exp\left\{-\frac{3\pi^{2}}{g^{2}(\A)}\beta-{\cal E}_{0}\beta
-\beta(F_{1}-F_{0}) \right\}.                                                 
\ee
In this expression, the renormalized gauge coupling constant is
\be                                           \label{5:23}
\frac{1}{g^{2}(\A)}=\frac{1}{g^{2}(\A_{0})}
-\frac{11}{12\pi^{2}}\ln\left(\frac{\A}{\A_{0}}\right).
\ee
Here we have returned to the dimensionful $\A$ and replaced
$g$ by $g(\A_{0})$, where $\A_{0}=1/\mm$. 
This expression agrees with the renormalization group flow,
such that it
does not depend on the scale $\A_{0}$ if $g(\A_{0})$ is
chosen to obey 
the Gell-Mann-Low equation. To fix the scale, we assume that
the value of $g(\A_{0})$ is determined by the typical energy
of the physical processes in the universe, that is by the physical
temperature $T(\A_{0})=1/\beta\A_{0}$. Then we use the QCD data
(see for example \cite{QCD})
\be                                                                            
                              \label{5:23j}
T(\A_{0})=100\ {\rm GeV},
\ \ \ \ \ \ \frac{g^{2}(\A_{0})}{4\pi}=0.12,
\ee
and assume that the weak coupling region extends up to some
$\A_{max}$.
One can choose $\A_{max}\sim 10\div 100\A_{0}$.

${\cal E}_{0}$ is  the contribution of the zero field
oscillations, that is, the Casimir energy,
\be                                             \label{5:23a}
{\cal E}_{0}=
{\rm PP}\zeta_{spat}(-\frac{1}{2})+\frac{11}{4}(\ln 2 -1).
\ee
It is worth noting that this quantity  can be computed exactly
in this case. The corresponding computation
itself presents some methodological interest and is given in the
Appendix.
The result is
$$
{\cal E}_{0}=\frac{5}{6}+\frac{11}{4}(\ln 2-1-\gamma)
+\int_{0}^{1}dz\sqrt{1-z^{2}}\times
$$
\be                                                \label{5:23b}
\times\int_{0}^{1}dt
\tan\left(\frac{\pi t}{2}\right)
\left\{(z^{2}+4)F(z,t)+(4z^{2}-2)G(z,t,\sqrt{2})+
9z^{2}G(z,t,\sqrt{3})\right\},
\ee
where
\be                                                \label{5:23c}
F(z,t)=t-\frac{\sinh(\pi z t)}{\sinh(\pi z)},\ \ \ \
G(z,t,q)=t-\frac{\sin(\pi qz t)}{\sin(\pi qz)}+
\frac{2}{\pi}\frac{\sin(\pi t)}{(1-q^{2}z^{2})}.
\ee
The numerical value is
\be                                               \label{5:23d}
{\cal E}_{0}=-1.084\ldots 
\ee
The contribution of the thermal degrees of freedom in
(\ref{5:20a}) is
$$
\beta (F_{1}-F_{0})=
4\ln(1-e^{-\beta})
+2\sum_{\sigma=0,1,2}\ \sum_{n=3}^{\infty}(n^{2}-\sigma^{2})
\ln(1-e^{-\beta\sqrt{n^{2}+\sigma^{2}-3}})-
$$
\be
-6\sum_{n=2}^{\infty}(n^{2}-1)
\ln(1-e^{-\beta n}),                                 \label{5:24}
\ee
and the remaining sums in this expression
can be evaluated numerically (see Fig.1). 

\begin{figure}
\epsfxsize=8cm
\centerline{\epsffile{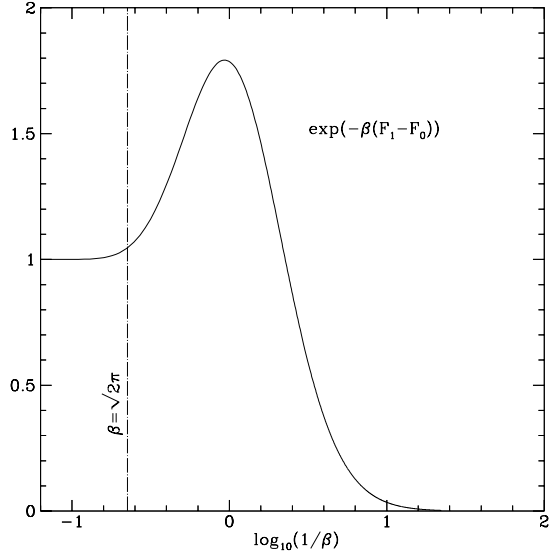}}
\caption{The thermal function $\exp(-\beta(F_{1}-F_{0}))$.}
\label{Fig.1}
\end{figure}

One can see that $F_{1}$ and $F_{0}$ are precisely the free
energies of the physical
oscillators (\ref{4:17}), (\ref{4:18}). 
Altogether Eqs.(\ref{5:20a})-(\ref{5:24}) provide the
desired solution
of the one-loop sphaleron transition problem. The numerical
curves of $\Gamma(1/\beta)$ evaluated according to these formulas
for several values of $\A$ are presented in Fig.2.

\begin{figure}
\epsfxsize=8cm
\centerline{\epsffile{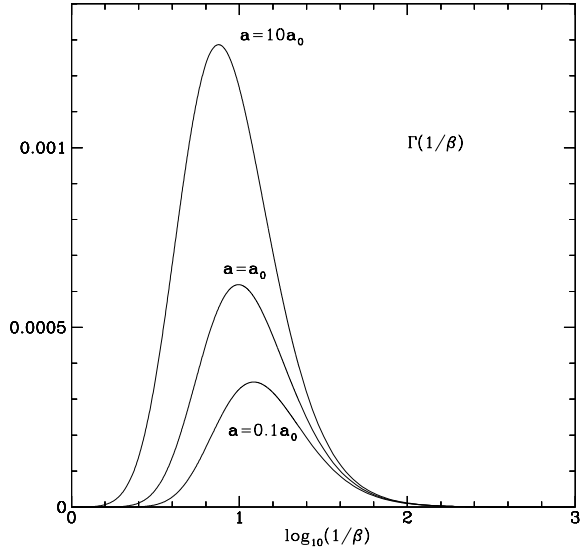}}
\caption{The sphaleron transition rate $\Gamma(1/\beta)$.}
\label{Fig.2}
\end{figure}

This solution makes sense under the following assumptions:
\be                                                   \label{5:24a}
\A\leq\A_{max},\ \ \ \ \ \
\frac{1}{\sqrt{2}\pi}<\frac{1}{\beta}\ll\frac{3\pi^{2}}{g^{2}(\A)}.
\ee
The first condition is the the weak coupling requirement.  
When the scale factor $\A$ is too large, the running coupling 
constant (\ref{5:23}) becomes big (confinement phase), and 
the effects of the strong coupling can completely change the 
semiclassical picture. That is why our solution can be trusted only
for small values of the size of the universe. The other condition 
in (\ref{5:24a}) requires that the thermal fluctuations are 
small compared to the classical sphaleron energy,  such that 
the perturbation theory is valid. 
Notice that each curve  in Fig.2 develops a maximum at some 
temperature   $1/\beta_{max}(\A)$; however, this value 
seems to be already beyond the scope of the approximation:
$3\pi^{2}/g^{2}(\A)\sim 2/\beta_{max}(\A)$. The subsequent
decrease of $\Gamma$ is presumably fictitious. Indeed, it is 
natural to expect that the transition rate is increasing with growing
temperature \cite{mottola}. Thus, our results can be trusted
at best only for $1/\beta<1/\beta_{max}(\A)$.

One can also find the high temperature limit for the solution
(but the upper bound in (\ref{5:24a}) is to be assumed).
To determine the asymptotic behavior of the free energy,
the procedure is the  following \cite{dowker}.
First, one returns to  the zeta function $\zeta_{\beta}(s)$ 
and replaces in (\ref{5:10})
the heat kernel $\Theta_{spat}(t)$ 
by its asymptotic expansion (\ref{5:12}). Then one takes the integral
over $t$, and the sum over $l$ reduces to the Riemann zeta function.
As a result, one arrives at the following asymptotic expansion
for small $\beta$ \cite{dowker}:
$$
\zeta_{\beta}'(0)=\frac{\pi^{2}}{45}\frac{C_{0}}{\beta^{3}}+
\frac{\zeta(3)}{2\pi\sqrt{\pi}}\frac{C_{1/2}}{\beta^{2}}+
\frac{C_{1}}{12\beta}-\frac{C_{3/2}}{4\pi\sqrt{\pi}}\ln\beta+
\frac{C_{2}}{8\pi^{2}}
\left(\gamma+\ln\frac{\beta}{4\pi}\right)\beta+
$$
\be
+\frac{1}{4\pi\sqrt{\pi}}\sum_{r\geq 1}C_{r+3/2}
\left(\frac{\beta}{2\pi}\right)^{2r}\zeta_{R}(2r)\Gamma(r)+
\zeta_{spat}'(0).                                      \label{5:25}
\ee
This gives the free energy (see Eq.(\ref{5:20}))
\be                                                  \label{5:25:1}
F=-\frac{1}{\beta}\zeta_{\beta}'(0)+
\frac{(1-\ln 2) C_{2}}{8\pi^{2}}-
{\rm PP}\zeta_{spat}(-\frac{1}{2}).
\ee
As an illustration we
apply this to the vacuum fluctuations alone.
The corresponding spatial zeta function is given by the last piece
in Eq.(\ref{5:18}):
$\zeta_{spat}^{0}(s)=3\zeta_{R}(2s-2)-3\zeta_{R}(2s)$,
its coefficients $C_{r}$ are computed in the Appendix
(one has $C_{2}=0$), and the result is
\be
F_{0}=-2\pi^{2}\left(\frac{\pi^{2}}{15\beta^{4}}-\frac{1}{2\beta^{2}}
-\frac{3}{2\pi^{2}}\frac{\ln\beta}{\beta}+
\frac{3}{4\pi^{2}\beta}+\frac{11}{80\pi^{2}}+O(\beta)
\right).
\ee
Here $2\pi^{2}$ is the volume of $S^{3}$ 
(remember that $\beta=1/ \A  T$), 
and the leading term
is just the free energy of a gas of noninteracting, massless
particles with $2\times 3$ polarization states.

Now, let us return to the full expression (\ref{5:18}) for
$\zeta_{spat}(s)$. 
The values of $C_{r}$ in this case are given by (\ref{5:14}).
Using (\ref{5:23a}), (\ref{5:25:1}) one obtains
\be                                                    \label{5:26}
-\beta(F_{1}-F_{0})=
4\ln\beta+\zeta_{spat}'(0)
+\left(\frac{11}{4}(\gamma+\ln\frac{\beta}{4\pi})+
{\cal E}_{0}\right)\beta+O(\beta^{2}).
\ee
The quantity $\zeta_{spat}'(0)$ is computed in the Appendix:
\be                                                     \label{5:27}
\kappa\equiv
\exp\{\zeta_{spat}'(0)\}=
\frac{2\sqrt{2}}{\pi^{3}}\sinh^{4}(\pi)
|\sin(\sqrt{2}\pi)|
\exp\left\{{\cal J}(1)-{\cal I}(\sqrt{2})-{\cal I}(\sqrt{3})\right\},
\ee
where
\be
{\cal J}(x)=\pi\int_{0}^{x}t^{2}\coth(\pi t)dt,\ \ \ \
{\cal I}(x)=\int_{0}^{x}\left(\frac{2t}{1-t^{2}}+
\pi t^{2}\cot(\pi t)\right)dt,
\ee
with the numerical value 
\be                                                     \label{5:28}
\kappa=1250.21\ldots 
\ee
(It is interesting to observe the large value of $\kappa$.
The corresponding quantity for the electroweak sphaleron
is suppressed by several orders of magnitude
\cite{carson}, \cite{baake}.)

Finally, one obtains (see Fig.3)
\be                                                    \label{5:29}
\exp\{-\beta(F_{1}-F_{0})\}=\kappa\beta^{4}\left(1+
\frac{11}{4}\beta\ln\beta\right)+O(\beta^{5})
\ \ \ {\rm as}\ \ \beta\rightarrow 0.
\ee

\begin{figure}
\epsfxsize=8cm
\centerline{\epsffile{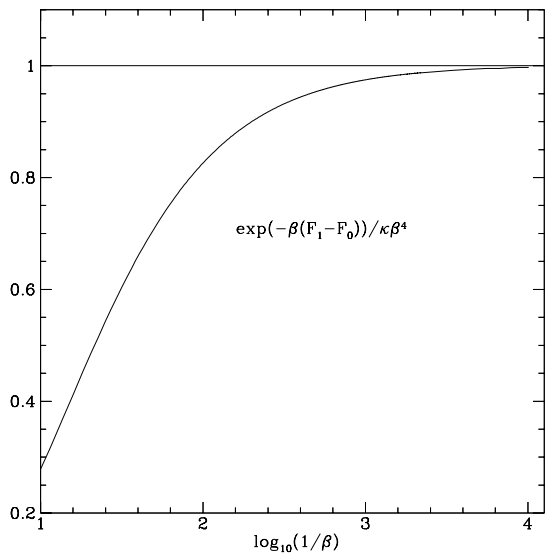}}
\caption{The ratio between the thermal function         
$\exp(-\beta(F_{1}-F_{0}))$ 
and its high temperature asymptote $1250.21\beta^{4}$.}
\label{Fig.3}
\end{figure}

As a result, we arrive at the following
expression for the sphaleron transition rate 
in the high temperature limit:
\be                                                    \label{5:30}
\Gamma(\beta)=\frac{\kappa\beta^{3}}{8\pi^{2}}
\exp\left\{-\frac{3\pi^{2}}{g^{2}(\A)}\beta\right\},
\ee
where we have neglected also the Casimir term.
As one can see from Fig.3, the high temperature approximation
can be reasonable for $1/\beta\geq 10^{3}$. On the other hand,
the temperature should be less then the sphaleron energy
$3\pi^{2}/g^{2}(\A)$. These two conditions imply that (\ref{5:30})
makes sense only for small $g(\A)$:
$10^{3}\leq 3\pi^{2}/g^{2}(\A)$, that is, for small $\A$.

\newp 
\section{Concluding remarks}
\setcounter{equation}{0}

In this paper we have obtained the exact solution
of the sphaleron transition problem for a
pure non-Abelian gauge field in a static Einstein universe.
This has been achieved by the straightforward
diagonalization of the one-loop
fluctuation operators with the subsequent
computation of the functional determinants 
in the zeta function regularization scheme.
To carry out this program, the following points
have been crucial: the high symmetry of the sphaleron solution
and the fact that the sphaleron configuration consists of the
gauge field alone. Actually, these properties of the model under
consideration make it somewhat similar to the instanton theory
\cite{thooft}, \cite{polyakov}, \cite{ore}.
It is worth noting that the solution obtained in this paper
is unique in the sense that
no other exact solutions of the sphaleron models
in $3+1$ spacetime dimensions are known.

Eqs.(\ref{5:20a})-(\ref{5:24}) and (\ref{5:30}) are our
principal results. They specify the number of transitions between
the neighboring topological sectors per unit conformal time $\eta$. 
It should be stressed that
such transitions do not lead to any violation
of chiral fermion number
unless the thermal equilibrium between the different
topological sectors is broken.
This can arise, for instance, when
a fermion asymmetry is present. Then one has
to introduce a small chemical potential $\mu$ for the fermions
\cite{shap}, \cite{arnold}. This favors those transitions which erase
the asymmetry. Specifically, let
$\Delta N=N_{F}-\bar{N}_{F}$ be the fermion number
of the universe, then 
\be                                                 \label{6:1}
\frac{d}{d\eta}\Delta N\approx -\frac{\mu}{T}\Gamma.
\ee
Notice that -- since $\Gamma$ defines the
number of  transitions in the whole universe -- $\Delta N$ refers 
to the whole universe as well, 
and not to the unit volume.   For one doublet of chiral fermions, 
standard thermodynamics gives
\be                                             \label{6:2}
N_{F}=\frac{V}{\pi^{2}}\int_{0}^{\infty}
\frac{p^{2}dp}{\exp\{\frac{p-\mu}{T}\}+1}=
\frac{3VT^{3}}{2\pi^{2}}\zeta_{R}(3)+
\frac{\mu}{6}VT^{2}+O(\mu^{2}),
\ee
where $V=2\pi^{2}\A^{3}$ is the volume of 3-space, such that 
$\Delta N=2\pi^{2}\A\mu/3\beta^{2}$.
Finally, passing to the physical
time $t=\A\eta$,
one obtains the fermion number diffusion rate 
\be                                                 \label{6:3}
\frac{1}{\Delta N}\frac{d}{dt}\Delta N=
-\frac{3\beta^{3}}{2\pi^{2}\A}\Gamma (\frac{1}{\beta}).
\ee

In fact, we do not specify the nature of the fields under consideration. 
The discussion of the possible applications of the
results obtained in this paper will be given separately.
At present we just mention where our results can be used.
Let us recall the typical values of the parameters $\A$, $\beta$ and
$T=1/\A\beta$. The range of $T$ is
restricted by the condition of the validity
of the semiclassical picture: $T\geq 1\div 10$ GeV (``deconfinement
phase''), whereas the metastability condition requires that 
the conformal temperature, $1/\beta$, is not too high
(see Eq.(\ref{5:24a})). The size of the universe,
$\A=1/T\beta$, therefore should not be too large compared to $1/T$.
Such conditions can be met in the context
of the finite volume QCD (the typical volume in that
case is $\A^{3}\sim 1$ fm$^{3}$; see \cite{baal} and references therein).
Another natural possibility relates to the pre-inflation cosmology.
In this case, the
gravitating Yang-Mills field can arise in the context
of a superstring theory \cite{string}. In fact,
the semiclassical sphaleron transition picture applies
also after inflation. 
However, the conformal temperature is enormously large then,
$1/\beta=\A T\sim s^{1/3}\sim 10^{28}$,
where $s$ is the total entropy
of the universe. Equivalently, one can say that the 
temperature $T$ is huge in comparison with the sphaleron barrier
$3\pi^{2}/g^{2}(\A)\A$, such that there is no suppression for 
transitions between the different topological sectors at all. 
Unfortunately, there are no reliable methods for
computing the transition rate $\Gamma$ in this limit. 
One can use a dimensional argumant to estimate 
that the rate related
to the unit physical 4-volume, $\Gamma/\A^{4}$, should be
proportional to $T^{4}$. Then $\Gamma(1/\beta)\sim 1/\beta^{4}$,
and (\ref{6:3}) yields
\be                                                 \label{6:4}
\frac{1}{\Delta N}\frac{d}{dt}\Delta N\sim T.
\ee
This agrees with the usual estimate for the fermion number
dissipation rate at very high temperatures \cite{kuzmin}.

\section*{Acknowledgments}

The discussions with Andreas Wipf are gratefully acknowledged.
The author also wishes to thank
Norbert Straumann, Jurg Fr\"ohlich, Daniel Wyler and Slava Mukhanov
for useful discussions, and Marcus Heusler for a careful reading
of the manuscript. 
This work was supported by the Swiss National Science Foundation
and by the Tomalla Foundation.

\section*{Appendix. Zeta function techniques}
\renewcommand{\theequation}{A.\arabic{equation}}
\setcounter{equation}{0}

In  Sections  {\bf A}--{\bf D} below
the detailed computation
of ${\rm PP}\zeta_{spat}(-1/2)$ and
$\zeta_{spat}'(0)$ is presented.
The basic idea is to
expand these quantities into certain series of the
Riemann zeta functions and  then to
use the appropriate summation formulae.
The asymptotic expansion for the heat kernel $\Theta_{spat}(t)$
is derived in Sec.{\bf E}. Section {\bf F}
contains the computation of $\zeta_{\beta}(0)$ and
$\zeta_{\beta}'(0)$ for a thermal system.

\noindent
{\bf A. Summation formulae}

\noindent
Consider the generating function for the Bernoulli polynomials
(see \cite{abram}, p.804)
\be                                                      \label{a1}
\frac{x e^{xt} }{e^{x}-1}=
\sum_{k=0}^{\infty}B_{k}(t)\frac{x^{k}}{k!},\ \ \ \ \ \ |x|<2\pi. 
\ee
Putting here $t=0$ one obtains
\be                                                      \label{a2}
\coth(x)=\frac{1}{x}\sum_{k=0}^{\infty}B_{2k}
\frac{(2x)^{2k}}{(2k)!},\ \ \ \ \ |x|<\pi,                 
\ee
where $B_{k}(0)=B_{k}$ are the Bernoulli numbers.
Using the relation between the Bernoulli numbers and
the Riemann zeta
function  (see \cite{abram}, p.807),
\be                                                      \label{a3}
B_{2k}=(-1)^{k+1}\frac{2(2k)!}{(2\pi)^{2k}}\zeta_{R}(2k),  
\ee
and considering the replacement $x\rightarrow i x$, one finds
\be                                                      \label{a4}
\coth(x)=-\frac{2}{x}\sum_{k=0}^{\infty}(-1)^{k}
\left(\frac{x}{\pi}\right)^{2k}\zeta_{R}(2k);\ \ \ \ \
\cot(x)=-\frac{2}{x}\sum_{k=0}^{\infty}
\left(\frac{x}{\pi}\right)^{2k}\zeta_{R}(2k).
\ee
The sums on the right hand sides converge only for $|x|<\pi$,
whereas the left hand sides are meramorphic functions on the whole
complex plane. One can therefore consider these relations
for any $x$ as a result of analytic continuation.

Let us now restrict ourselves to the real values of $x$. 
Integrating both sides of (\ref{a4}) one obtains
\be                                                    \label{a5}
\sum_{k=1}^{\infty}\frac{(-1)^{k}}{k}
\left(\frac{x}{\pi}\right)^{2k}\zeta_{R}(2k)=
\ln\frac{x}{\sinh(x)};\ \ \ \ \ \
\sum_{k=1}^{\infty}\frac{1}{k}
\left(\frac{x}{\pi}\right)^{2k}\zeta_{R}(2k)=
\ln\frac{x}{\sin(x)}.                                    
\ee
In the second of these formulae one should assume that $x<\pi$,
unless the integration rule for the poles of $\cot(x)$ is specified
(see below).

Multiplying (\ref{a4}) by $x^{2}$ and integrating from
zero to $x$ one obtains
\be
\sum_{k=1}^{\infty}\frac{(-1)^{k}}{k}
\left(\frac{x}{\pi}\right)^{2k}\zeta_{R}(2k-2)=
{\cal J}\left(\frac{x}{\pi}\right),\ \ \ 
\sum_{k=1}^{\infty}\frac{1}{k}
\left(\frac{x}{\pi}\right)^{2k}\zeta_{R}(2k-2)=
\tilde{{\cal J}}\left(\frac{x}{\pi}\right),  \label{a5:1}
\ee
where
\be                                                 \label{a5:2}
{\cal J}(x)=\pi\int_{0}^{x}t^{2}\coth(\pi t)dt,\ ({\rm any}\ x);\ \ \
\tilde{{\cal J}}(x)=-\pi\int_{0}^{x}t^{2}\cot(\pi t)dt
\ \ (x<\pi).                                              
\ee
Eqs.(\ref{a5})-(\ref{a5:2}) will be used below.
To proceed further, we return for a moment to Eq.(\ref{a1}).
Consider the equation which is obtained from (\ref{a1}) under the
replacement $x\rightarrow -x$. Taking the difference of the two
equations  one obtains
\be                                                   \label{a5:3}
\frac{e^{xt} }{e^{x}-1}+ \frac{e^{-xt} }{e^{-x}-1}=
2\sum_{k=0}^{\infty}B_{2k+1}(t)\frac{x^{2k}}{(2k+1)!}.
\ee
Using  the explicit form of the $k=0$ term on the right hand side,
$2B_{1}(t)=2t-1$, and utilizing also the following
relation (see \cite{abram}, p.807):
\be                                                   \label{a5:4}
\zeta_{R}(2k+1)=(-1)^{k+1}\frac{(2\pi)^{2k+1}}{2(2k+1)!}
\int_{0}^{1}B_{2k+1}(t)\cot(\pi t)dt,
\ee
one finds
\be                                             \label{a5:5}
\sum_{k=1}^{\infty}(-1)^{k}\left(\frac{x}{2\pi}\right)^{2k}
\zeta_{R}(2k+1)=-\frac{\pi}{2}\int_{0}^{1}\left(1-2t+
\frac{e^{xt} }{e^{x}-1}+ \frac{e^{-xt} }{e^{-x}-1}\right)
\cot(\pi t)dt.
\ee
Replacing here $x\rightarrow 2x$ and $t\rightarrow 2t-1$,
and considering also $x\rightarrow i x$
one arrives at
\be                                             \label{a5:6}
\sum_{k=1}^{\infty}(-1)^{k}\left(\frac{x}{\pi}\right)^{2k}
\zeta_{R}(2k+1)=-\frac{\pi}{2}I(x),\ \ \ \ \
\sum_{k=1}^{\infty}\left(\frac{x}{\pi}\right)^{2k}
\zeta_{R}(2k+1)=-\frac{\pi}{2}\tilde{I}(x),
\ee
where
\be                                             \label{a5:6:1}
I(x)=\int_{0}^{1}\left(t-
\frac{\sinh(xt)}{\sinh(x)}\right)\tan\left(\frac{\pi t}{2}\right)dt,
\ \ \ \ \ \tilde{I}(x)=
\int_{0}^{1}\left(t-
\frac{\sin(xt)}{\sin(x)}\right)\tan\left(\frac{\pi t}{2}\right)dt.
\ee
Next, one deduces from (\ref{a5:6}) that
\be                                             \label{a5:7}
\sum_{k=1}^{\infty}(-1)^{k}\nu^{2k} z^{2k}
\zeta_{R}(2k+1)=
-\frac{\pi}{2}I(\pi\nu z).
\ee
Using 
\be                                               \label{a5:8}
\frac{\Gamma(k+1/2)}{\Gamma(k+2)}=\frac{4}{\sqrt{\pi}}
\int_{0}^{1}d z z^{2k}\sqrt{1- z^{2}},\ \ \
\frac{\Gamma(k+1/2)}{\Gamma(k+3)}=\frac{8}{3\sqrt{\pi}}
\int_{0}^{1}d z z^{2k}\left(1- z^{2}\right)^{3/2},\ \ \
\ee
we finally obtain the following formulae:
$$
\sum_{k=1}^{\infty}(-1)^{k}\nu^{2k}\frac{\Gamma(k+1/2)}
{\Gamma(k+2)}
\zeta_{R}(2k+1)=-2\sqrt{\pi}\int_{0}^{1}d z\sqrt{1- z^{2}}
I(\pi\nu z),
$$
\be                                                   \label{a5:9}
\sum_{k=1}^{\infty}(-1)^{k}\nu^{2k}\frac{\Gamma(k+1/2)}
{\Gamma(k+3)}
\zeta_{R}(2k+1)=-\frac{4}{3}\sqrt{\pi}\int_{0}^{1}d z
\left(1- z^{2}\right)^{3/2}
I(\pi\nu z),
\ee
together with the corresponding two relations obtained by
$\nu\rightarrow i\nu$, $I(\pi\nu z)\rightarrow \tilde{I}(\pi\nu z)$.
Here the integrals can be considered
for arbitrary values of $\nu$, which defines the analytic extension
of the series. 

\vspace{1 mm}

\noindent
{\bf B. The basic zeta function relation}

\noindent
Consider the following zeta function
\be                                                  \label{a6}
\zeta_{\nu}^{0}(s)=\sum_{n=1}^{\infty}
\frac{1}{(n^{2}+\nu^{2})^{s}}=
\frac{1}{\Gamma(s)}\int_{0}^{\infty}t^{s-1}\sum_{n=1}^{\infty}
\exp\left\{-t(n^{2}+\nu^{2}) \right\}dt.
\ee
In order to express this function in terms of the Riemann zeta function, we
expand $\exp(-t\nu^{2})=\sum_{k}(-1)^{k}\nu^{2k}t^{k}/k!$, and 
perform the integration over $t$. Then the
sum over $n$ gives the Riemann zeta function, such that
\be                                                  \label{a7}
\zeta_{\nu}^{0}(s)=\zeta_{R}(2s)+
\sum_{k=1}^{\infty}
\frac{(-1)^{k}}{k!}\nu^{2k}
\frac{\Gamma(k+s)}{\Gamma(s)}\zeta_{R}(2k+2s).
\ee
This relation will turn out to be especially handy. 
Taking the pole of the gamma function at $s=0$ into account, 
using Eq.(\ref{a5})
and utilizing $\zeta_{R}(0)=-\frac{1}{2}$,
$\zeta_{R}'(0)=-\frac{1}{2}\ln 2\pi$, one finds
\be                                                   \label{a8}
\zeta_{\nu}^{0}(0)=-\frac{1}{2},\ \ \
\frac{d}{ds}\zeta_{\nu}^{0}(0)=-\ln(2\pi)+\sum_{k=1}^{\infty}
\frac{(-1)^{k}}{k}\nu^{2k}\zeta_{R}(2k)=
\ln\frac{\nu}{2\sinh(\pi\nu)}.
\ee
For a harmonical oscillator, for instance, one has
\be                                                  \label{a8:1}
\zeta(s)=\sum_{l=1}^{\infty}
\left\{\left(\frac{2\pi l}{\beta}\right)^{2}+\omega^{2}\right\}^{-s}=
\left(\frac{\beta}{2\pi}\right)^{2s}\zeta_{\beta\omega/2\pi}^{0}(s)
\ \ \Rightarrow\ \
\frac{d}{ds}\zeta(0)=
\ln\frac{\omega}{2\sinh(\beta\omega/2)},
\ee
which gives rise to the formula (\ref{5:8}) in the main text
when $\omega=i\sqrt{2}$.

We shall also use another zeta function
\be                                                    \label{a8:2}
\zeta_{\nu}^{2}(s)=\sum_{n=1}^{\infty}
\frac{n^{2}}{(n^{2}+\nu^{2})^{s}}=
\zeta_{R}(2s-2)+
\sum_{k=1}^{\infty}
\frac{(-1)^{k}}{k!}\nu^{2k}
\frac{\Gamma(k+s)}{\Gamma(s)}\zeta_{R}(2k+2s-2).
\ee
Using Eqs.(\ref{a5:1}) and (\ref{a5:2}) one obtains
\be                                                   \label{a8:3}
\zeta_{\nu}^{2}(0)=0,\ \ \ \ \
\frac{d}{ds}\zeta^{2}_{\nu}(0)=2\zeta_{R}'(-2)+{\cal J}(\nu),
\ee
where ${\cal J}(\nu)$ is defined by (\ref{a5:2}).

\vspace{1 mm}


\noindent
{\bf C. Evaluation of ${\rm PP}\zeta_{spat}(-\frac{1}{2})$}

\noindent
We consider the spatial zeta function
\be                                                   \label{d1}
\zeta_{spat}(s)=2+\sum_{\sigma=0,1,2}\ \sum_{n=3}^{\infty}
\frac{n^{2}-\sigma^{2}}{(n^{2}+\sigma^{2}-3)^{s}}-
3\sum_{n=2}^{\infty}\frac{n^{2}-1}{(n^{2})^{s}},        
\ee
which  can be represented in the form
\be                                                    \label{d2}
\zeta_{spat}(s)={\cal A}(s)+{\cal B}(s),
\ee
where
\be                                                    \label{d3}
{\cal A}(s)=2+3\zeta_{R}(2s)-3\zeta_{R}(2s-2)+\sum_{\nu^{2}}
(\nu^{2}-1)\left(4+\nu^{2}\right)^{-s},
\ee
and
\be                                                    \label{d4}
{\cal B}(s)=\sum_{\nu^{2}}\left\{
\zeta^{0}_{\nu}(s-1)-(1+\nu^{2})^{1-s}\right\}-\sum_{\nu^{2}}
(3+2\nu^{2})\left\{\zeta^{0}_{\nu}(s)-(1+\nu^{2})^{-s}
\right\}.
\ee
Here $\zeta^{0}_{\nu}(s)$ is defined by Eq.(\ref{a6}), and
$\nu^{2}=\sigma^{2}-3=1,-2,-3$. In ${\cal A}(s)$ one can simply
put $s=-1/2$, which yields
${\rm PP}{\cal A}(-1/2)={\cal A}(-1/2)=-91/40-3\sqrt{2}$.

Let us now consider ${\cal B}(s-1/2)$ in the limit $s\rightarrow 0$.
We first analyze $\zeta^{0}_{\nu}(s-1/2)$.
Using Eq.(\ref{a7}) we obtain
\be                                                      \label{d5}
\zeta_{\nu}^{0}(s-\frac{1}{2})=\zeta_{R}(2s-1)-
\nu^{2}\frac{\Gamma(s+\frac{1}{2})}
{\Gamma(s-\frac{1}{2})}\zeta_{R}(2s+1)+
\sum_{k=2}^{\infty}
\frac{(-1)^{k}}{k!}\nu^{2k}
\frac{\Gamma(k+s-1/2)}{\Gamma(s-\frac{1}{2})}\zeta_{R}(2k+2s-1).
\ee
Let us now consider the limit where $s$ tends to zero. Then the second
term on the right hand side diverges due to the pole of the
Riemann zeta function,
\be                                                   \label{d6}
\zeta_{R}(1+s)=\frac{1}{s}+\gamma+O(s^{2}),
\ee
where $\gamma$ is the Euler constant. The remaining terms in
(\ref{d5}) are all finite. The principal part is
\be                                                   \label{d7}
{\rm PP}\zeta_{\nu}^{0}(-\frac{1}{2})=
\lim_{s\rightarrow 0}\frac{d}{ds}
\left(s\zeta_{\nu}^{0}(s-\frac{1}{2})\right).
\ee
Using $\zeta_{R}(-1)=-1/12$, $\Gamma(-1/2)=-2\sqrt{\pi}$,
and replacing $k\rightarrow k+1$
in the sum entering Eq.(\ref{d5}), one obtains
\be                                                  \label{d8}
{\rm PP}\zeta_{\nu}^{0}(-\frac{1}{2})=-\frac{1}{12}-
\frac{1}{2}\nu^{2}(1-\gamma)+\frac{\nu^{2}}{2\sqrt{\pi}}
\sum_{k=1}^{\infty}(-1)^{k}\nu^{2k}\frac{\Gamma(k+1/2)}
{\Gamma(k+2)}
\zeta_{R}(2k+1).
\ee
Finally, taking into account Eqs.(\ref{a5:9}), one arrives at
\be                                                    \label{d9}
{\rm PP}\zeta_{\nu}^{0}(-\frac{1}{2})=-\frac{1}{12}-
\frac{1}{2}\nu^{2}(1-\gamma)-\nu^{2}
\int_{0}^{1}dz\sqrt{1-z^{2}}I(\pi\nu z),
\ee
where $I(x)$ is defined by Eq.(\ref{a5:6:1}). Similarly, one
obtains
\be                                                    \label{d10}
{\rm PP}\zeta_{\nu}^{0}(-\frac{3}{2})=\frac{1}{120}-
\frac{1}{8}\nu^{2}-
\frac{1}{2}\nu^{4}(1-\frac{3}{4}\gamma)-\nu^{4}
\int_{0}^{1}dz\left(1-z^{2}\right)^{3/2}I(\pi\nu z).
\ee

The next step is to insert these relations  in Eq.(\ref{d4})
and to compute the sum over $\nu^{2}=1, -2, -3$. The $\nu^{2}=1$ case
presents no problems, whereas the negative values of
$\nu^{2}$ should
be treated with some care. Let us pass in (\ref{d9}) to negative
$\nu^{2}$, $\nu\rightarrow iq$, where 
$q=|q|$. For $q<1$ one obtains
\be
{\rm PP}\zeta_{\nu}^{0}(-\frac{1}{2})\ \rightarrow\                            
                    \label{d11}
{\rm PP}\zeta_{iq}^{0}(-\frac{1}{2})=-\frac{1}{12}+
\frac{1}{2}q^{2}(1-\gamma)+q^{2}
\int_{0}^{1}dz\sqrt{1-z^{2}}\tilde{I}(\pi qz),
\ee
where $\tilde{I}(x)$ is defined by Eq.(\ref{a5:6:1}).
Next we consider the following integral representation for 
$\sqrt{1-q^{2}}$ which is valid for $0<q<1$:
\be                                                   \label{d12}
\sqrt{1-q^{2}}=1-\frac{2q^{2}}{\pi}\int_{0}^{1}
dz\sqrt{1-z^{2}}\int_{0}^{1}
\frac{\sin(\pi t)}{1-q^{2}z^{2}}
\tan\left(\frac{\pi t}{2}\right) dt.
\ee
This implies
$$
{\rm PP}\zeta_{iq}^{0}(-1/2)-\sqrt{1-q^{2}}=
-1-\frac{1}{12}+
\frac{1}{2}q^{2}(1-\gamma)+
$$
\be                                                    \label{d13}
+q^{2}
\int_{0}^{1}dz\sqrt{1-z^{2}}
\int_{0}^{1}\left(t-
\frac{\sin(\pi z  qt)}
{\sin(\pi z q)}+\frac{2}{\pi}\frac{\sin(\pi t)}{1-q^{2}z^{2}}
\right)\tan\left(\frac{\pi t}{2}\right)dt.
\ee
Here we can safely extend the range of $q$ from
$0<q<1$ to $0<q<2$, thus taking the values
$q=\sqrt{2}$,$\sqrt{3}$ into account,
that is, $\nu^{2}=-2$,$-3$. This allows us
to compute the second sum over $\nu^{2}$ in the formula
(\ref{d4}) for ${\rm PP}{\cal B}(-1/2)$. The first sum  can be
done with a similar rearrangement, using the
formula
\be                                                     \label{d14}
(1-q^{2})^{3/2}=1-\frac{3}{2} q^{2}+
\frac{2q^{4}}{\pi}\int_{0}^{1}
dz\left(1-z^{2}\right)^{3/2}\int_{0}^{1}
\frac{\sin(\pi t)}{1-q^{2}z^{2}}
\tan\left(\frac{\pi t}{2}\right) dt.
\ee
Finally, collecting everything together, we arrive at
\be
{\rm PP}\zeta_{spat}(-\frac{1}{2})=\frac{5}{6}-\frac{11}{4}
\gamma+
5{\cal R}_{1}(1)-{\cal R}_{3}(1)+
+2{\cal P}_{1}(\sqrt{2})-4{\cal P}_{3}(\sqrt{2})+
9{\cal P}_{1}(\sqrt{3})-9{\cal P}_{3}(\sqrt{3});
\ee
where
$$
{\cal R}_{m}(\nu)=
\int_{0}^{1}dz \left(1-z^{2}\right)^{m/2}
\int_{0}^{1}\left(t-
\frac{\sinh(\pi z\nu t)}
{\sinh(\pi z\nu)}\right)\tan\left(\frac{\pi t}{2}\right)dt,
$$
\be
{\cal P}_{m}(q)=
\int_{0}^{1}dz \left(1-z^{2}\right)^{m/2}
\int_{0}^{1}\left(t-
\frac{\sin(\pi z q t)}
{\sin(\pi z q)}+\frac{2}{\pi}\frac{\sin(\pi t)}{(1-q^{2}z^{2})}
\right)\tan\left(\frac{\pi t}{2}\right)dt.
\ee

\vspace{1 mm}

\noindent
{\bf D. Evaluation of $\zeta_{spat}'(0)$}

\noindent
The procedure in this case is similar to that of the previous section.
One again starts from Eqs.(\ref{d2})-(\ref{d4}), but now,
using (\ref{a6}) and (\ref{a8:2}), it is convenient to
represent the function
${\cal B}(s)$ in the following equivalent form:
\be                                                    \label{e1}
{\cal B}(s)=\sum_{\nu^{2}}\left\{
\zeta^{2}_{\nu}(s)-(1+\nu^{2})^{-s}\right\}-\sum_{\nu^{2}}
(3+\nu^{2})\left\{\zeta^{0}_{\nu}(s)-(1+\nu^{2})^{-s}
\right\}.
\ee
The next steps are straightforward. Using (\ref{a8}) and
(\ref{a8:3}) one obtains
\be                                                    \label{e2}
{\cal B}'(0)=\sum_{\nu^{2}}\left\{2\zeta_{R}'(-2)+{\cal J}(\nu)
+\ln(1+\nu^{2})\right\}-\sum_{\nu^{2}}
(3+\nu^{2})\ln\frac{\nu(1+\nu^{2})}{2\sinh(\pi\nu)},
\ee
where ${\cal J}(\nu)$ is defined by (\ref{a5:2}), and $\nu^{2}=1$,
$-2$, $-3$. For negative $\nu^{2}=-q^{2}<0$ one again needs
some minor modifications. No changes are needed 
for the second sum in (\ref{e2}):
\be                                                    \label{e3}
\ln\frac{\nu(1+\nu^{2})}{2\sinh(\pi\nu)}\ \ \rightarrow\ \ \
\ln\frac{q(1-q^{2})}{2\sin(\pi q)},
\ee
where we can put  $q=\sqrt{2}$, $\sqrt{3}$. In the first sum
one has
\be
{\cal J}(\nu)+\ln(1+\nu^{2})\ \rightarrow\ \
\tilde{{\cal J}}(q)+\ln(1-q^{2})=
-\pi\int_{0}^{q}t^{2}\cot(\pi t)dt-
\int_{0}^{q}\frac{2tdt}{1-t^{2}}\equiv -{\cal I}(q),
\ee
where 
\be                                                       \label{e4}
{\cal I}(q)=\int_{0}^{q}\left(\frac{2t}{1-t^{2}}+
\pi t^{2}\cot(\pi t)\right)dt,
\ee
which can also be extended to the values $q=\sqrt{2}$, $\sqrt{3}$.
For ${\cal A}'(0)$ one has
\be                                                     \label{e5}
{\cal A}'(0)=-3\ln(2\pi)-6\zeta_{R}'(-2)+3\ln 2,
\ee
which finally gives
\be                                                        \label{e6}
\zeta_{spat}'(0)=\ln\left(\frac{2\sqrt{2}}{\pi^{3}}\sinh^{4}(\pi)
|\sin(\sqrt{2}\pi)|\right)+{\cal J}(1)-
{\cal I}(\sqrt{2})-{\cal I}(\sqrt{3}),
\ee
with ${\cal J}(x)$ and ${\cal I}(x)$ being defined in
Eqs.(\ref{a5:2}) and (\ref{e4}), respectively.


\vspace{1 mm}

\noindent
{\bf E. Asymptotic expansion of $\Theta_{spat}(t)$}

\noindent
Let us consider the following function
\be
\Theta(i,j,k|t)=\sum_{n=i}^{\infty}(n^{2}-j)e^{-(n^{2}+k)t}=
-e^{-kt}\left(\frac{d}{dt}+j\right)
\sum_{n=i}^{\infty}e^{-n^{2}t}.                          \label{a9}
\ee
We want to find its asymptotic expansion for small $t$,
\be
\Theta(i,j,k|t)\sim\frac{1}{(4\pi t)^{3/2}}
\sum_{r=0,1/2,1,\ldots}C_{r}t^{r}.                      \label{a10}
\ee
Such an expansion can be obtained with the use of the theta
function identity \cite{hille}
\be                                                   \label{a11}
\sum_{n=-\infty}^{\infty}\exp(-t n^{2})=\sqrt{\frac{\pi}{t}}
\sum_{n=-\infty}^{\infty}\exp\left\{-\frac{\pi^{2}}{t}n^{2}\right\},
\ee
which implies that
\be
\sum_{n=i}^{\infty}\exp(-t n^{2})\sim
\frac{1}{2}\sqrt{\frac{\pi}{t}}-\frac{1}{2}-
\sum_{n=1}^{i-1}\exp(-t n^{2}),                      \label{a12}
\ee
since for $t\rightarrow 0$
all terms with $\exp\{-\pi^{2}n^{2}/t\}$, $n\neq 0$
vanish faster than any power of $t$ and
can therefore be omitted. Inserting this into (\ref{a9})
and expanding the remaining exponents we find
\be
\Theta(i,j,k|t)\sim
\frac{\sqrt{\pi}}{4t^{3/2}}-
\frac{\sqrt{\pi}}{4t^{1/2}}(k+2j)+
j/2+\sum_{n=1}^{i-1}(j-n^{2})+
\frac{\sqrt{\pi}}{8}k(k+4j)t^{1/2}+O(t).           \label{a13}
\ee
The coefficients $C_{r}$ for $r\leq 2$ are therefore  
\be
C_{0}=2\pi^{2},\ C_{1/2}=0,\ C_{1}=-2\pi^{2}(k+2j),\
C_{3/2}=8\pi\sqrt{\pi}
(\frac{j}{2}+\sum_{n=1}^{i-1}(j-n^{2})),\
C_{2}=\pi^{2}k(k+4j).                               \label{a14}
\ee
The heat kernel (\ref{5:11}) in the main text
can be represented as
\be
\Theta_{spat}(t)=2e^{-t}+\sum_{\sigma=0,1,2}
\Theta(3,\sigma^{2},\sigma^{2}-3|t)-3\Theta(2,1,0|t),  \label{a15}
\ee
which finally gives rise to Eq.(\ref{5:14}).

\vspace{1 mm}

\vspace{1 mm}

\noindent
{\bf F. Evaluation of $\zeta_{\beta}(0)$ and $\zeta_{\beta}'(0)$
\cite{kirsten}}

\noindent
Consider the thermal zeta function for an arbitrary system of
harmonic oscillators with positive energies
\be
\zeta_{\beta}(s)=
\frac{1}{\Gamma(s)}\int_{0}^{\infty}t^{s-1}\sum_{l=-\infty}^{\infty}
\exp\left\{-\left(\frac{2\pi l}{\beta}\right)^{2}t\right\}
\Theta(t)dt,                                             \label{a16}
\ee
where
\be                                                      \label{a17}
\Theta(t)=\sum_{\omega}\exp\{-\omega^{2}t\},\ \ \ \omega^{2}>0.
\ee
Transforming the sum over $l$ in
(\ref{a16}) with the use of the theta function identity (\ref{a11})
and using the integral representation for the Kelvin functions
\be                                                     \label{a18}            
                                  \label{5:16}
K_{\nu}(z)=\frac{1}{2}\left(\frac{z}{2}\right)^{\nu}
\int_{0}^{\infty}dt\
t^{-\nu-1}\exp\left\{-t-\frac{z^{2}}{4t}\right\},
\ee
one obtains \cite{kirsten}
\be                                                   \label{a18:1}
\zeta_{\beta}(s)=\frac{\beta}{2\sqrt{\pi}}
\frac{1}{\Gamma(s)}
Y_{spat}(s-\frac{1}{2})+
\frac{2\beta}{\sqrt{\pi}\Gamma(s)}\sum_{l=1}^{\infty}
\sum_{\omega}\left(\frac{\beta l}{2\omega}\right)^{s-1/2}
K_{1/2-s}(\beta l\omega),                               
\ee
where
\be                                                    \label{a19}
Y_{spat}(s)=\int_{0}^{\infty}dt\ t^{s-1}\Theta(t)=
\zeta_{spat}(s)\Gamma(s).                                
\ee
This function has the following pole structure \cite{voros}
\be                                                    \label{a20}
Y_{spat}(s)=\frac{1}{(4\pi)^{3/2}}
\sum_{r}\frac{C_{r}}{s+r-3/2}+f(s),                      
\ee
where $C_{r}$ are defined by the asymptotic
expansion of $\Theta(t)$, and 
$f(r)$ is an entire analytic function of $s$. This relation implies
\be                                                    \label{a21}
Y_{spat}(s-\frac{1}{2})=\frac{C_{2}}{(4\pi)^{3/2}}\frac{1}{s}+
{\rm PP}Y_{spat}(-\frac{1}{2}).                          
\ee
Taking Eq.(\ref{a19}) and the
properties of the gamma function into account,
\be                                                   \label{a22}
\frac{1}{\Gamma(s)}=s+\gamma s^{2}+O(s^{3}),\ \ \ \ \
\Gamma(-\frac{1}{2}+s)=
-2\sqrt{\pi}\left\{1+(-\gamma+2-2\ln 2)s\right\}+O(s^{3}),
\ee
one therefore obtains
\be
\frac{1}{2\sqrt{\pi}\Gamma(s)}Y_{spat}(s-\frac{1}{2})=
\frac{C_{2}}{16\pi^{2}}+
\left\{\frac{(1-\ln 2) C_{2}}{8\pi^{2}}-
{\rm PP}\zeta_{spat}(-\frac{1}{2})\right\}s+O(s^{2}).   \label{a23}
\ee
Finally, using
\be
K_{1/2}(z)=\sqrt{\frac{\pi}{2z}}e^{-z},\ \ \ \ \ \  \ \ \ 
\sum_{l=1}^{\infty}\frac{1}{l}e^{-lz}=-\ln(1-e^{-z}),  \label{a24}
\ee
one arrives at
\be                                                \label{a25}
\zeta_{\beta}(s)=\frac{C_{2}\beta}{16\pi^{2}}+
\left\{\left(\frac{(1-\ln 2) C_{2}}{8\pi^{2}}-
{\rm PP}\zeta_{spat}(-\frac{1}{2})\right)\beta-
2\sum_{\omega}\ln\left(1-e^{-\beta\omega}\right)\right\}s+O(s^{2}),
\ee
which gives rise to the formula (\ref{5:20}) in the main text.

\newp


\begin{thebibliography}{99}
\bibitem{rebbi} R.Jackiw, C.Rebbi,
{\it Phys.Rev.Lett.} {\bf 37} (1976) 172.

\bibitem{thooft} G.'t Hooft, {\it Phys.Rev.} {\bf D 14} (1976) 3432.
 
\bibitem{Manton} N.S.Manton, {\it Phys.Rev.} {\bf D 28} (1983) 2019;

F.R.Klinkhamer, N.S.Manton, {\it Phys.Rev.} {\bf D 30} (1984) 2212.

\bibitem{kuzmin} V.A.Kuzmin, V.A.Rubakov, M.E.Shaposhnikov,
{\it Phys.Lett.} {\bf B 155} (1985) 36.

\bibitem{arnold}
P.Arnold, L.McLerran, {\it Phys.Rev.} {\bf D 36} (1987) 581;
{\it Phys.Rev.} {\bf D 37} (1988) 1020.

\bibitem{shap} A.I.Bochkarev, M.E.Shaposhnikov,
{\it Mod.Phys.Lett.} {\bf A 2} (1987) 417.





\bibitem{carson} L.Carson, X.Li, L.McLerran, R.T.Wang,
{\it Phys.Rev.} {\bf D 42} (1990) 2127.

\bibitem{baake} J.Baake, S.Junker, {\it Phys.Rev.}
{\bf D 49} (1994) 2055;
{\it Phys.Rev.} {\bf D 50} (1994) 4227.

\bibitem{diak} D.Diakonov, M.Polyakov, P.Sieber, J.Schaldach, K.Goeke,
{\it Phys.Rev.} {\bf D 53} (1996) 3366;
{\it Phys.Rev.} {\bf D 49} (1994) 6864.

\bibitem{Forgacs} P.Forgacs, Z.Horvath,
{\it Phys.Lett.} {\bf B 138} (1984) 397;

N.S.Manton, T.N.Samols, {\it Phys.Lett.} {\bf B 207} (1988) 179;



\bibitem{bochkarev} A.I.Bochkarev, M.E.Shaposhnikov,
{\it Mod.Phys.Lett.} {\bf A 2} (1987) 991.



\bibitem{mottola} E.Mottola, A.Wipf, {\it Phys.Rev.}
{\bf D 39} (1989) 588.

\bibitem{BK} R. Bartnik, J. McKinnon, {\it Phys. Rev. Lett.}
{\bf 61} (1988) 141.

\bibitem{10} D.V. Gal'tsov, M.S. Volkov, {\it Phys. Lett.}
{\bf B 273} (1991) 255. 

M.S. Volkov, {\it Phys. Lett.} {\bf B 328} (1994) 89; 
{\it Phys. Lett.} {\bf B 334} (1994) 40.

\bibitem{12} M.S.Volkov, O.Brodbeck, G.Lavrelashvili,
N.Straumann,
{\it Phys. Lett.} {\bf B 349} (1995) 438.

\bibitem{Hosotani} Y.Hosotani, {\it Phys.Lett.}
{\bf B 147} (1984) 44.

\bibitem{henneaux} M.Henneaux, {\it Journ.Math.Phys.}
{\bf 23} (1982) 830.

\bibitem{cosmEYM} J.Cervero, L.Jacobs, {\it Phys.Lett.}
{\bf B 78} (1978) 427;


Y.Verbin, A.Davidson, {\it Phys.Lett.} {\bf B 229} (1989) 364;

A.Hosoya, W.Ogura, {\it Phys.Lett.} {\bf B 225} (1989) 117;

D.V.Gal'tsov, M.S.Volkov, {\it Phys.Lett.} {\bf B 256} (1991) 17;

O.Bertolami, J.M.Mourao, R.F.Picken, I.P.Volobujev,
{\it Int.Journ.Mod.Phys.} {\bf A 6} (1991) 4149.

\bibitem{baal} P. van Baal, N.D.Hari Dass, {\it Nucl.Phys.}
{\bf B 385} (1992) 185;

A.V.Smilga, {\it Nucl.Phys.} {\bf B 459} (1996) 263.

\bibitem{cosmsphal} G.W.Gibbons, A.R.Steif, 
{\it Phys.Lett.} {\bf B 320} (1993) 243;
{\it Phys.Lett.} {\bf B 346} (1995) 255;

\bibitem{ding} S.Ding, {\it Phys.Rev.}
{\bf D 50} (1994) 3755.

\bibitem{alfaro} V.De Alfaro, S.Fubini, G.Furlan, {\it Phys.Lett.}
{\bf B 65} (1976) 163;

M.L\"uscher, {\it Phys.Lett.} {\bf B 70} (1977) 321;

B.Schechter, {\it Phys.Rev.} {\bf D 16} (1977) 3015.

\bibitem{sing} E.Farli, V.V.Khoze, R.Singleton, Jr.,
{\it Phys.Rev.} {\bf D 47} (1993) 5551;

C.Rebbi, R.Singleton, Jr., BUHEP-95-33, hep-ph/9601260.

\bibitem{callan} C.G.Callan, S.Coleman,
{\it Phys.Rev.} {\bf D 16} (1977) 1762.


\bibitem{polyakov} A.M.Polyakov,
{\it Nucl.Phys.} {\bf B 120} (1977) 429;

A.A.Belavin, A.M.Polyakov, {\it Nucl.Phys.} {\bf B 123} (1977) 429.

\bibitem{ore} F.R.Ore, {\it Phys.Rev.} {\bf D 16} (1977) 2577;

S.Chadha, P.Di Vecchia, A.D'Adda, F.Nicodemi,
{\it Phys.Lett.} {\bf B 72} (1977) 103.

\bibitem{osborn} H.Osborn, {\it Ann.Phys.} {\bf 135} (1981) 373.

\bibitem{langer} L.S.Langer, {\it Ann.Phys.} {\bf 41} (1967) 108;
{\it Ann.Phys.} {\bf 54} (1969) 258;

I.Affleck, {\it Phys.Rev.Lett} {\bf 46} (1981) 388.

\bibitem{QCD} G.Altarelli, CERN-TH/96-05.



\bibitem{zeta} J.S.Dowker, R.Crichley, {\it Phys.Rev.}
{\bf D 13} (1976) 3224;

S.W.Hawking, {\it Comm.Math.Phys.} {\bf 55} (1977) 133.

\bibitem{dowker} J.S.Dowker, G.Kennedy, {\it J.Phys.}
{\bf A 11} (1978) 895.

\bibitem{connor} D.J.O'Connor, B.L.Hu, T.C.Shen,
{\it Phys.Lett.} {\bf B 130} (1983) 31.

\bibitem{voros} A.Voros, {\it Comm.Math.Phys.} {\bf 110} (1987) 439.

\bibitem{wipf}
S.Blau, M.Visser, A.Wipf, {\it Nucl.Phys.} {\bf B 310} (1988) 163;
{\it Phys.Lett.} {\bf B 209} (1988) 209;

C.Wiessendanger, A.Wipf, {\it Ann.Phys.} {\bf 233} (1994) 125.

\bibitem{kirsten} K.Kirsten, E.Elizalde, 
hep-th/9508068. 

\bibitem{hille} E.Hille, {\it Analytic Function Theory, vol II},
Ginn, Boston, 1962.


\bibitem{abram} M.Abramowitz, I.Stegun, {\it Handbook of
Mathematical Functions}, Dover, New York, 1965. 

\bibitem{string} M.B.Green, J.H.Schwarz, E.Witten,
{\it Superstring Theory}, Cambridge University Press, 1987.

\end{thebibliography}
\end{document}